\documentclass[11pt, a4paper]{article}
\usepackage{graphicx}

\usepackage[all]{xy}
\usepackage[latin1]{inputenc}
\usepackage{amsmath}
\usepackage{amsfonts}
\usepackage{amssymb}
\usepackage{vmargin}
\usepackage{graphicx}

\begin{document}

\title{Numerical solution for a non-Fickian diffusion in a periodic potential}



\author{Ad\'erito Ara\'ujo$^{a,}$\footnote{This work was partially supported by the research project UTAustin/MAT/066/2008}
, Amal K. Das$^{b}$, Cid\'alia Neves$^{a,c,*}$, Erc\'\i lia Sousa$^{a,*,}$
\footnote{Corresponding author: ecs@mat.uc.pt} \\ \\
 {\small $^{a}$CMUC, Department of Mathematics, University of Coimbra, 3001-454 Coimbra, Portugal} \\ 
{\small $^{b}$Department of Physics, Dalhousie University, Halifax, Nova Scotia B3H 3J5, Canada }\\
{\small $^{c}$ISCAC, Polytechnic Institute of Coimbra, 3040-316 Coimbra, Portugal}}

\maketitle

\begin{abstract}
Numerical solutions of a non-Fickian diffusion equation belonging to a
hyperbolic type are presented in one space dimension. The Brownian
particle modelled by this diffusion equation is subjected to a symmetric
periodic potential whose spatial shape can be varied by a single 
parameter. We consider a numerical method which consists of applying
Laplace transform in time; we then obtain an elliptic diffusion equation
which is discretized using a finite difference method. We analyze some
aspects of the convergence of the method. Numerical results for
particle density, flux and mean-square-displacement (covering both
inertial and diffusive regimes) are  presented. 
\end{abstract}

{\it keywords}: numerical methods, Laplace transform, telegraph equation, periodic 

potential, non-Fickian diffusion

\section{Introduction}

In this paper we shall present numerical solutions of a non-Fickian
diffusion equation in the presence of a symmetric periodic potential in
one space dimension. Let us briefly recall that the Fickian diffusion
equation in the presence of a potential $V(\xi)$ reads
\begin{equation}
\frac{\partial n}{\partial \tau}(\xi,\tau)
= D\frac{\partial^2 n}{\partial \xi^2}(\xi,\tau)
+\frac{1}{m\gamma}\frac{\partial}{\partial \xi}\left[
  \frac{dV}{d\xi}(\xi)n(\xi,\tau)\right],
\label{diffusion}
\end{equation}
where $\gamma$ is a friction parameter and $D=k_BT/m\gamma$ is the diffusion
coefficient, $m$ being the mass of the Brownian particle whose overdamped
(diffusive) dynamics is well described by (\ref{diffusion}),
$k_B$ is the Boltzmann's constant and $T$ the temperature of the
fluid.

The equation of our study is
\begin{equation}
\frac{1}{\gamma}\frac{\partial^2 n}{\partial \tau^2}(\xi,\tau)+\frac{\partial n}{\partial \tau}(\xi,\tau)
 =  D\frac{\partial^2 n}{\partial \xi^2}(\xi,\tau)
+\frac{1}{m\gamma}\frac{\partial}{\partial \xi}\left[
  \frac{dV}{d\xi}(\xi)n(\xi,\tau)\right].
\label{telegraph}
\end{equation}
Both equations, (\ref{diffusion}) and (\ref{telegraph}) can be derived from an
underlying kinetic equation e.g. the phase-space Kramers equation \cite{das1991}
\begin{equation}
\frac{\partial f}{\partial \tau}+\frac{p}{m}\frac{\partial f}{\partial \xi}
-\frac{dV}{d\xi}\frac{\partial f}{\partial p} =\gamma\frac{\partial}{\partial
  p}(pf)
+mk_BT\gamma\frac{\partial^2 f}{\partial p^2},
\label{kramers}
\end{equation}
where $f(\xi,p,\tau)$ is the probability density function for the position
component $\xi$ and momentum component $p$  of the Brownian particle.

Equation (\ref{telegraph}) in the absence of a potential field is sometimes
referred to the telegrapher equation although we shall call it a non-Fickian
diffusion equation. We refer to \cite{das1991} for a derivation of
(\ref{telegraph}) from (\ref{kramers}). It may be noted that for times larger
than $1/\gamma$ , the first term on the left hand side of (\ref{telegraph})
can be neglected and the Fickian regime is regained.  Equation
(\ref{diffusion}) in the absence of a potential field leads to the well known
result for the mean square displacement \cite{fm2002}
\begin{equation}
< \xi^2(\tau) > = 2D\tau.
\label{mean}
\end{equation}
In the presence of a flexible symmetric potential, it was shown in \cite{das1991}
that $<\xi^2(\tau)>$ does not necessarily behave linearly with time. Equation
(\ref{telegraph})
retains some short time inertial behaviour of a Brownian particle and
at long time results in a diffusive behaviour.
The velocity $v=d\xi/d\tau$ of
a Brownian particle is not well defined in the diffusive regime for
which (\ref{diffusion}) is applicable. Since (\ref{telegraph}) is applicable in an inertial
regime, the velocity can be calculated with (\ref{telegraph}). Quite recently the
instantaneous velocity of a Brownian particle has been experimentally
investigated \cite{huang2011,li2010,pus2011}. This provides an additional motivation for studying
(\ref{telegraph}).
There is also  a recent paper \cite{barbero2010} which models transport of ions in
insulating media through a non-Fickian diffusion equation of the type
discussed in our work. In \cite{barbero2010} the non-Fickian diffusion equation is
referred to as a hyperbolic diffusion equation .

To solve our problem we consider a numerical method  based on space discretization and 
time Laplace transform. The latter is suitable for long times and also for
solutions that are not necessarily smooth in time. It may be noted that 
iterative methods in time, including implicit methods such as the Crank-
Nicolson \cite{cn47}, which allows a choice of large time steps, usually take too long to
compute the solution.

The paper is organized as follows. In section 2 we present the model
problem in dimensionless variables. In section 3 we describe a
numerical method based on the time Laplace transform which is suitable
for long time integration and also for solutions which are not very
smooth.  In section 4 the convergence properties of the algorithm are
studied. In section 5 we present the behaviour of the solution to the
non-Fickian diffusion equation, the flux and the mean square
displacement. We conclude the paper, in section 6, with a summary and
outlook.

\section{The model and physical quantities}

In our studies we consider three quantities of physical interest. These are the
particle density \ $ n(\xi,\tau) $, \ the \ current \ density \ (flux) \ $ j(\xi,\tau) $  and \ the \ mean
square \ displacement \newline $<\xi^2(\tau)>$. The current
density is not normally studied. However, since we are dealing with a non-Fickian
diffusion equation we have decided to consider $j(\xi,\tau)$ as well. For the Fickian
case and in the absence of any potential, $j(\xi,\tau)$ is related to $n(\xi,\tau)$
through $j =-D (\partial n/\partial \xi)$. This is not so in the non-Fickian case for which the
relation between $j(\xi,\tau)$ and $n(\xi,\tau)$ is more involved.

Let us consider the non-Fickian diffusion equations for particle density
and the flux
\begin{eqnarray}
\frac{1}{\gamma}\frac{\partial^2 n}{\partial \tau^2}+\frac{\partial n}{\partial
  \tau}
& = & -\frac{1}{m\gamma}\frac{\partial}{\partial \xi}(Pn)+D\frac{\partial^2 n}{\partial \xi^2},\label{telegraph1}\\
j + \frac{1}{\gamma} \frac{\partial j}{\partial \tau} & = & -D\frac{\partial
  n}{\partial \xi} - \frac{1}{m\gamma}P n, \label{flux}
\end{eqnarray}
with $n(\xi,\tau)$ as  the density of the Brownian particles. $P$ is the force acting on
the particle due to the potential field $V$, i.e. 
$$P=-\frac{d V}{d\xi}.$$

\begin{figure}[h]
\centerline{
\includegraphics[height=50mm,width=70mm]{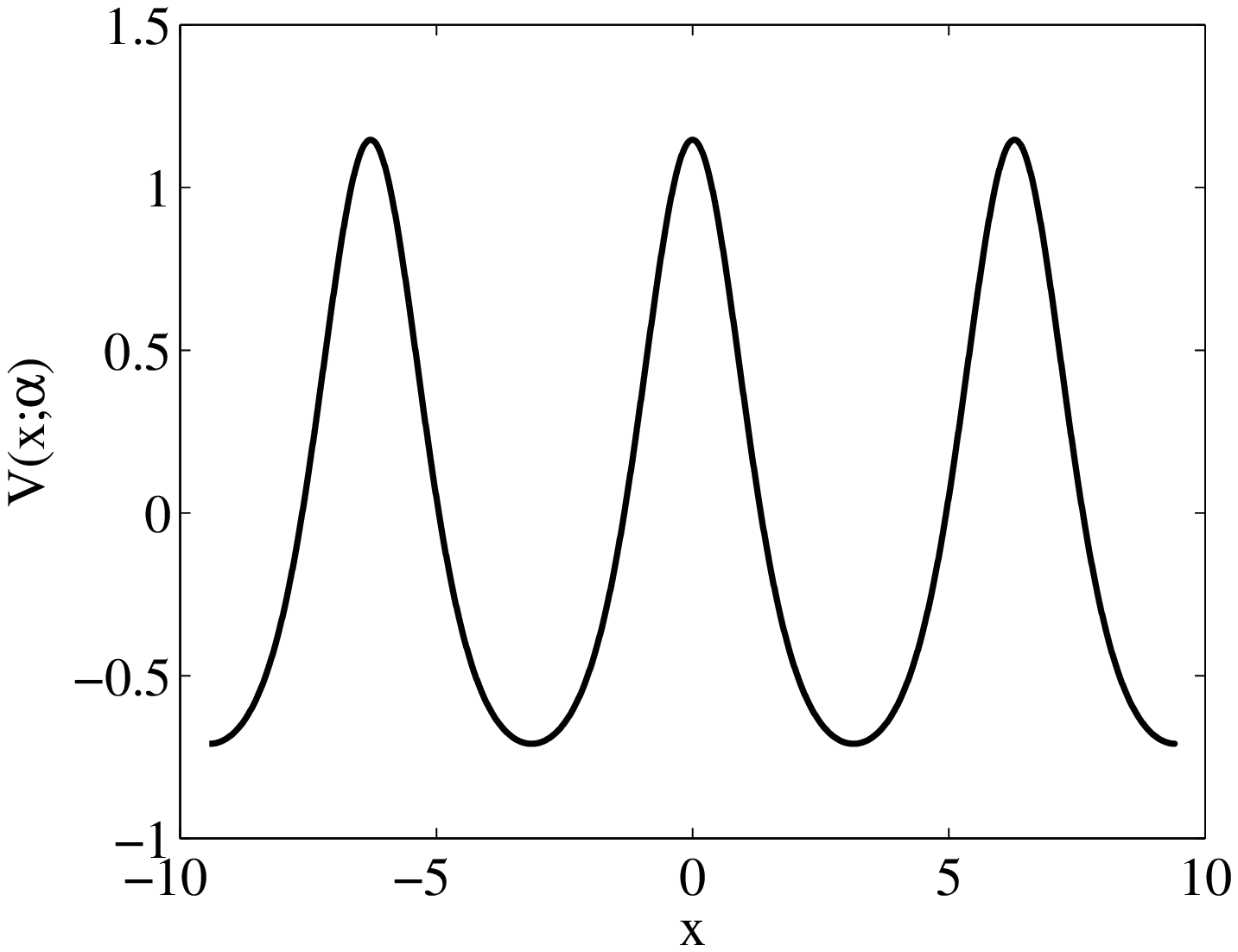}\qquad
\includegraphics[height=50mm,width=70mm]{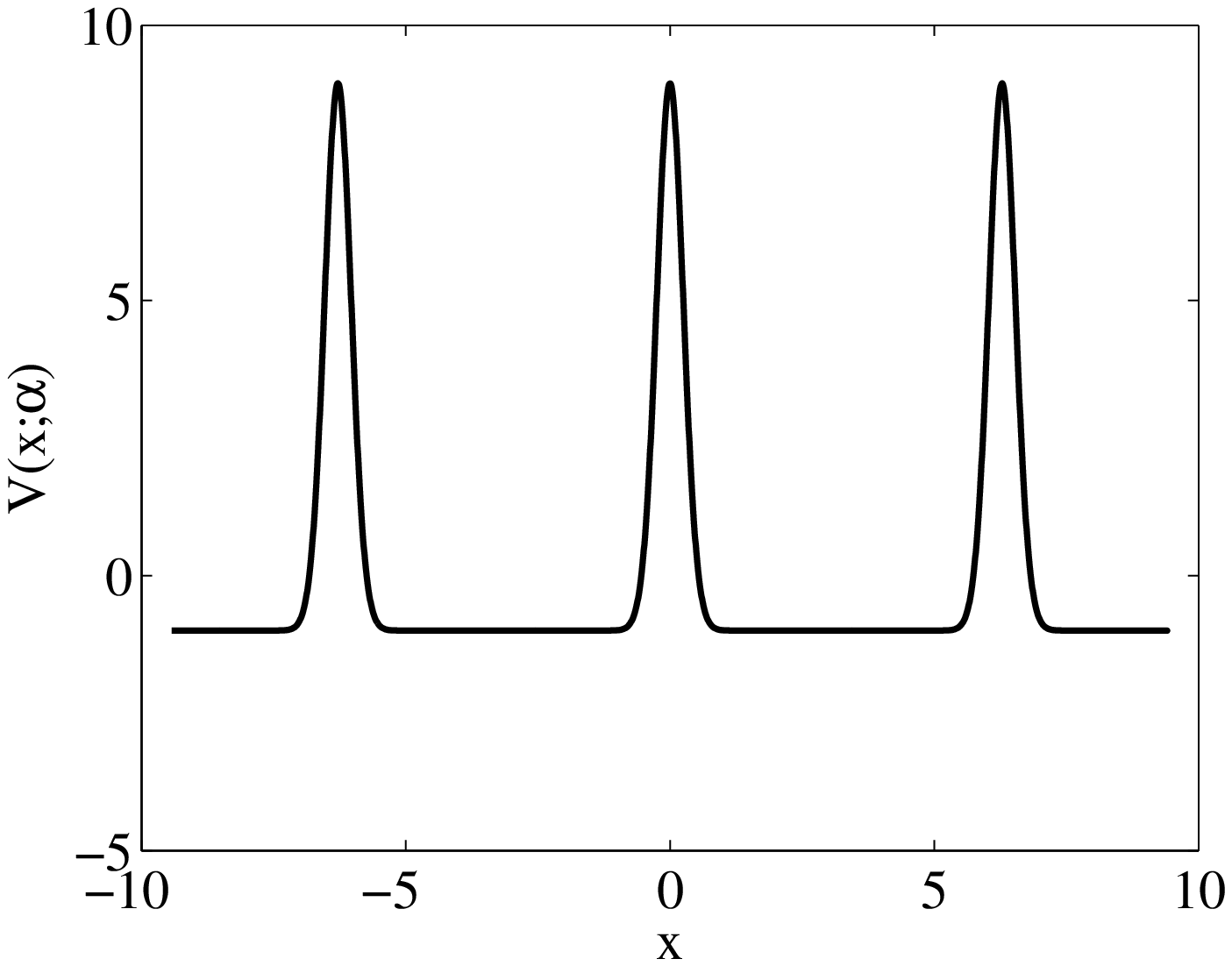}}
\caption{Potential field $V(x;\alpha)$. Left figure: $\alpha =1$; Right figure: $\alpha =16$.}
\label{potential1and16}
\end{figure}

We consider a symmetric periodic potential field, as previously studied in \cite{branka2000},
\cite{das1991} and \cite{magnasco1993}. It reads
\begin{equation}\label{valpha}
V(\xi;\alpha)=\frac{1}{J_0(\bold i\alpha)}e^{\alpha\cos{\xi}}-1,
\end{equation}
where $J_0$ is the Bessel function of the first kind and zero order and $\bold i$ is the imaginary number.
In order to illustrate the flexible form of this single-parameter potential we have plotted, Figure \ref{potential1and16},
the potential (\ref{valpha}),  for  two values of the parameter,
$\alpha=1$ and $\alpha=16$.

Our model consists of equations (\ref{telegraph1}) and (\ref{flux}), and the
potential field $V(\xi;\alpha)$ given by (\ref{valpha}). 
For later purpose we introduce the following dimensionless parameters
\begin{equation}\label{parameters}n=\frac{n}{n_0},\qquad x=\frac{\xi}{\sqrt{D/\gamma}},\qquad t=\tau\gamma,
\end{equation}
where $n_0$ is a reference particle density (concentration). The dimensionless
forms of (\ref{telegraph}) and (\ref{flux})  can be written as
\begin{eqnarray}
\frac{\partial^2 n}{\partial t^2}+\frac{\partial n}{\partial t}
& = & -\frac{\partial}{\partial x}(P n)+\frac{\partial^2 n}{\partial
  x^2},\label{telegraph_dim}
\\
j +  \frac{\partial j}{\partial t} & = & -\sqrt{D}\sqrt{\gamma}\frac{\partial
  n}{\partial x} + \gamma P(x) n, \label{flux_dim}
\end{eqnarray}
with
\begin{equation}\label{potential_dim}
P(x)=-\frac{1}{m\sqrt{D\gamma^3}}\frac{dV}{dx}.
\end{equation}

\section{Numerical method}
We consider equations (\ref{telegraph_dim}) and (\ref{flux_dim}) with 
the following initial conditions
\begin{eqnarray}
n\left(x,0\right) &=&\frac{1}{L \sqrt{\pi}}e^{-x^2/L^2},{\ \ \ \ }\frac{\partial n }{\partial t}\left( x,0\right) = 0,
\\
j\left(x,0\right) &=&\frac{1}{L
  \sqrt{\pi}}e^{-x^2/L^2}\left(\frac{\sqrt{D}\sqrt{\gamma}}{L^2}2x+\gamma
  P(x)\right),
\end{eqnarray}
where
$$
P(x)=-\frac{1}{m\sqrt{D\gamma^3}}\frac{dV}{dx}, \qquad V(x;\alpha)=\frac{1}{J_0(\bf{i}\alpha)}e^{\alpha\cos{x}}-1.
$$
The boundary conditions are given by
\begin{equation}
\lim_{x\rightarrow \infty} n(x,t)=0, \quad \lim_{x\rightarrow -\infty}n(x,t)=0
\end{equation}
and
\begin{equation}
\lim_{x\rightarrow \infty} j(x,t)=0, \quad \lim_{x\rightarrow -\infty}j(x,t)=0.
\end{equation}

In this section we describe a numerical method to solve
the problem (\ref{telegraph_dim})--(\ref{flux_dim}). Our approach can be
separated in three steps. First, we apply the Laplace transform to
(\ref{telegraph_dim})--(\ref{flux_dim}) in order to remove the time dependent
terms and we obtain an ordinary differential equation in $x$ that
also depends on the Laplace transform parameter $s$. Secondly, we
solve the ordinary differential equation obtained using a finite difference
scheme.
Lastly, using a
numerical inverse Laplace transform algorithm
we obtain the final approximate solution.

\subsection{Spatial discretization}

Our numerical method is facilitated if we apply time Laplace transform to
equation  (\ref{telegraph_dim}) and obtain the ordinary differential equation
\begin{equation}\label{laplace}
\frac{d^{2}\widetilde{n}}{dx^{2}}-\lambda_s\widetilde{n}-\frac{%
d }{dx}\left( P\widetilde{n}\right) =-(1+s)n(x,0),
\end{equation}
where $\lambda_s=s^{2}+s$, $s$ is a complex variable and
$\widetilde{n}$ is the Laplace transform of $n$
defined by
$$
\widetilde{n}(x,s) = \int_{0}^{\infty} {\rm e}^{-st} n(x,t) dt.
$$

Now, assume we have a space discretization $x_i = a+i \Delta x, \ i=0,
\dots, N$. Let
 $\widetilde{\eta}_i(s), \ i=0, \dots, N$ represent the
approximation of
 $\widetilde{n}\left(x_i, s\right)$ in the Laplace
transform domain.
 The outflow boundary is such that $\tilde{\eta}_N(s)=0$,
for all $s$
 and $N$ sufficiently large, which is according to the physical
boundary condition.

To derive the numerical method we consider central
differences to approximate the first derivative and the second
derivative of equation (\ref{laplace}). We obtain, for a fixed $s$, the finite difference scheme given by
\begin{equation}
\frac{\widetilde{\eta}_{i-1}(s)-2\widetilde{\eta}_i(s)+\widetilde{\eta}_{i+1}(s)}{\Delta
x^2}-\lambda_s\widetilde{\eta}_i(s)-\frac{P_{i+1}\widetilde{\eta}_{i+1}(s)-P_{i-1}\widetilde{\eta}_{i-1}(s)}{2\Delta
x}=-(1+s)n(x_i,0),
\label{diffmethod}
\end{equation}
for  $i=1,\ldots,N-1$, where $P_{i}=P(x_{i})$.

Therefore, we obtain the linear
system
\begin{equation}
K\left(s\right)\widetilde{\bf \eta}\left(s\right)=\widetilde{b}
\left(s\right),\label{system}
\end{equation}
where $K(s)=[K_{i,j}(s)]$ is a band matrix of size $N-1 \times N-1$
with bandwidth three and $ \widetilde{\bf \eta}\left(s\right)=[\widetilde{\eta}_1\left(s\right),\ldots,
\widetilde{\eta}_{N-1}\left(s\right)]^T$.
The matrix $K(s)$ has entries of the form
\begin{eqnarray}
K_{i,i-1}(s) & = & \frac{1}{\Delta x^2}+\frac{P_{i-1}}{2\Delta x}, \nonumber\\
 K_{i,i}(s) & = & -\frac{2}{\Delta x^2}-\lambda_s,\nonumber\\
 K_{i,i+1}(s) & = & \frac{1}{\Delta x^2}-\frac{P_{i+1}}{2\Delta x},
\label{kis}
\end{eqnarray}
and
$\widetilde{b}\left(s\right)$ contains boundary conditions, being
represented by
\begin{equation}
\widetilde{b}\left(s\right)=\left[%
\begin{array}{c}
  -(1+s)n(x_1,0) \\
  -(1+s)n(x_2,0) \\
  \vdots \\
  -(1+s)n(x_{N-2},0)\\
  -(1+s)n(x_{N-1},0) \\
\end{array}%
\right]+
\left[%
\begin{array}{c}
  -K_{1,0}(s)\widetilde{\eta}_0(s) \\
 0 \\
  \vdots \\
 0 \\
  -K_{N-1,N}(s)\widetilde{\eta}_{N}(s) \\
\end{array}%
\right].
\end{equation}

To compute the flux, we apply the Laplace transform to equation
(\ref{flux_dim}),
that is,
\begin{equation}\label{laplaceflux}
(1+s)\widetilde{j}  =  -\sqrt{D}\sqrt{\gamma}\frac{d\widetilde{n}}{dx}
+\gamma P(x)\widetilde{n} +j(x,0),
\end{equation}
where $\widetilde{j}$ is the Laplace transform of the flux $j$.
The last step is to determine an approximate solution $\eta_{i}\left(t\right)$ and $j_{i}(t)$  of $n(x_i,t)$ and $j(x_i,t)$
respectively, which is obtained
from $\widetilde{\eta}_{i}\left( s\right)$
and $\widetilde{j}_{i}\left(s\right)$ by using a
Laplace inversion numerical method.

\subsection{Laplace transform inversion}

In this section, we determine an {approximate} solution
$\eta_{i}\left(t\right)$ from $\widetilde{\eta}_{i}\left(s\right)$ by
using a Laplace inversion numerical method.  For the sake of clarity
we omit the index $i$, denoting $\widetilde{\eta}_{i}\left( s\right)$
by $\widetilde{\eta}\left( s\right)$.

A formally exact inverse Laplace transform of $\widetilde{\eta}\left(
s\right)$  into $\overline{\eta}\left( t\right)$ is given through the Bromwich integral \cite{mar1999}
\begin{equation}
\overline{\eta}\left( t\right) = \frac{1}{2\pi {\bf i}}\int_{\beta - {\bf i}\infty }^{\beta+{\bf i}\infty }
{\rm e}^{st}  \widetilde{\eta}\left( s\right)ds,
\end{equation}
where $\beta$ is such that the contour of integration is to the right-hand side of any singularity of 
$\widetilde{\eta}\left( s\right)$. 
However, for a numerical evaluation the above integral is first
transformed to an equivalent form
\begin{equation}
\overline{\eta}\left( t\right) =\frac{1}{\pi }{\rm e}^{ \beta t}
\int_{0}^{\infty }\mbox{Re}\left\{ \widetilde{\eta}\left( s\right) {\rm
e}^{{\bf{i}}\omega t} \right\} \hspace{0.01in}d\omega ,
\label{lint}
\end{equation}
where $s=\beta +{\bf{i}}\omega $ \cite{aba1995,mar1999,ANS2008}. 
The integral is now evaluated through the trapezoidal rule \cite{aba1995, crump1976}, with step
size $ \pi /T$, and we obtain
\begin{equation}
\overline{\eta}(t) =\frac{1}{T}{\rm e}^{\beta t} \left[ \frac{ \widetilde{\eta}%
\left( \beta \right) }{2}+\sum_{k=1}^{\infty } \mbox{Re}\left\{ \widetilde{\eta%
}\left( \beta +\frac{{\bf{i}}k\pi }{T}\right) {\rm e}^{\frac{{\bf{i}}k\pi t}{T}}
\right\} \right] -E_{T}, \label{series}
\end{equation}
for $0<t<2T$ and where $E_{T}$ is the discretization error. It is
known that the infinite series in this equation converges very
slowly. To accelerate the convergence, we apply the
quotient-difference
algorithm, proposed in \cite{ahn2003}, and also used in \cite{ANS2008}, to calculate the series in (\ref%
{series}) by the rational approximation in the form of a continued
fraction. Under some conditions we can always associate a continued
fraction to a given power series.

We denote $v\left( z\right)$ the continued fraction
\begin{equation}
v\left( z\right) =d_{0}/\left( 1+d_{1}z/\left( 1+d_{2}z/\left(
1+\cdots \right) \right) \right)  \label{continued0}
\end{equation}
associated to the power series in (\ref{series}). For $z={\rm e}^{{\bf{i}}\pi t/T}$,
\begin{equation}
v\left( z\right) = \frac{%
\widetilde{\eta}\left( \beta \right) }{2}+\sum_{k=1}^{\infty } \widetilde{\eta}%
\left( \beta +\frac{{\bf{i}}k\pi }{T}\right) z^k,\label{continued}
\end{equation}
{and the coefficients $d_i$'s of (\ref{continued0}) are obtained by recurrence
relations from the coefficients $\widetilde{\eta}
\left( \beta +\frac{{\bf{i}}k\pi }{T}\right)$.}

Let the $M$-th partial fraction be denoted by $v(z,M)$.
Therefore
\[
v\left( z\right) =v\left( z,M\right) +E_{F}^{M},
\]
where $E_{F}^{M}$ is the truncation error.
Then
\[
\overline{\eta}\left( t\right) =\frac{1}{T}{\rm e}^{\beta t}
\mbox{Re}\left\{ v\left( z,M\right) +E_{F}^{M}\right\} - E_{T}.
\]
The approximation for $\overline{\eta}\left( t\right) $ is denoted by
$\eta(t)$ and given by
\[
\eta\left( t\right) =\frac{1}{T}{\rm e}^{\beta t} \mbox{Re} \left\{
v\left( z,M\right) \right\} \mbox{.}
\]

\section{Convergence of the numerical method}

In this section we discuss the convergence of the numerical method
chosen to compute an approximate solution to equation (\ref{telegraph_dim}).
Let us denote
by $\widetilde{E}_S$ the error associated to the spatial
discretization, that is,
\begin{equation}
\widetilde{n}(x_i,s) = \widetilde{\eta}_{i}(s) +
\widetilde{E}_S(x_i,s). \label{errors}
\end{equation}
The next errors come from the numerical inversion of Laplace
transform, where the Laplace inverse transform of
$\widetilde{\eta}_{i}(s)$ is, as described in the previous section, the
solution
\begin{equation}
\overline{\eta}_{i}\left(t\right) =\frac{1}{T}{\rm e}^{\beta t}
\mbox{Re}\left\{ v\left( z,M_i\right) +E_{F}^{M}(x_i,t)\right\} -
E_{T}(x_i,t), \label{errorinvl}
\end{equation}
where $E_T$ is the error associated with the trapezoidal
approximation and $E_{F}^{M}$\ is the truncation error associated to
the continued fraction. Note that for each $x_i$ the algorithm
chooses a $M_i$ and therefore for each $x_i$ we have a different
value of the approximation of the continued fraction, $v\left(
z,M_i\right)$. Therefore from (\ref{errors})--(\ref{errorinvl}) we
have
\[
{n}(x_i,t) =  \frac{1}{T}{\rm e}^{\beta t} \mbox{Re}\left\{ v\left(
z,M_i\right) +E_{F}^{M}(x_i,t)\right\} - E_{T}(x_i,t) +
{E}_S(x_i,t),
\]
where ${E}_S(x_i,t)$ is the inverse Laplace transform of the error
$\widetilde{E}_S(x_i,s)$.

\

{\it Approximation errors $E_T$ and $E_F$}

\

The error $E_T$ that comes from the integral approximation using the
trapezoidal rule, according to Crump \cite{crump1976}, is
\[
E_T= \sum_{n=1}^{\infty} {\rm e}^{-2n\beta T} n(x_i,2nT+t).
\]
Assume now that our function is bounded such as $|n(x_i,t)|\leq {\rm
e}^{\sigma t}$, for all $x_i$. Therefore the error can be bounded by
\[
E_T \leq {\rm e}^{\sigma t} \sum_{n=1}^{\infty} {\rm
e}^{-2nT(\beta-\sigma) } = \frac{{\rm e}^{\sigma t}}{{\rm
e}^{2T(\beta-\sigma)}-1}, \quad 0 < t < 2T.
\]%
It follows that by choosing $\beta$ sufficiently larger than
$\sigma$, we can make $E_T$ as small as desired. For practical
purposes and in order to choose a convenient $\beta$ we use the
inequality which bounds the error
\[
E_T \leq {\rm e}^{\sigma t-2T(\beta-\sigma)}.
\]
If we want to have the bound $E_T \leq b_T$ then by applying the
logarithm in both sides of the previous inequality we have
\[
\beta \geq \sigma\frac{2T+t}{2T} -\frac{1}{2T}\ln\left({b_T}%
\right).
\]
Assuming $\sigma \geq 0 $ we can write
\[
\beta \geq \sigma -\frac{\ln\left(b_T\right)}{2T}.
\]
In our example we consider $\sigma=0$.  In practice the trapezoidal
error $E_T$ is controlled by the parameter $\beta$ we choose.

The second error, $E_{F}^{M}$, comes from the approximation of the
continued fraction given by (\ref{continued}). This error is
controlled by imposing a tolerance $TOL$ such as
\[
\left\vert v\left( z,M\right) -v\left( z,M-1\right) \right\vert
<TOL,
\]
in order to get the approximation $\eta_{i}(t) $ given by
\[
\eta_{i}(t) = \frac{1}{T}{\rm e}^{\beta t}
\mbox{Re}\{v\left(z,M_i\right)\},
\]
where $M_i$ changes according to which $x_i$ we are considering.

\begin{figure}[h]
\centerline{
\includegraphics[height=50mm,width=70mm]{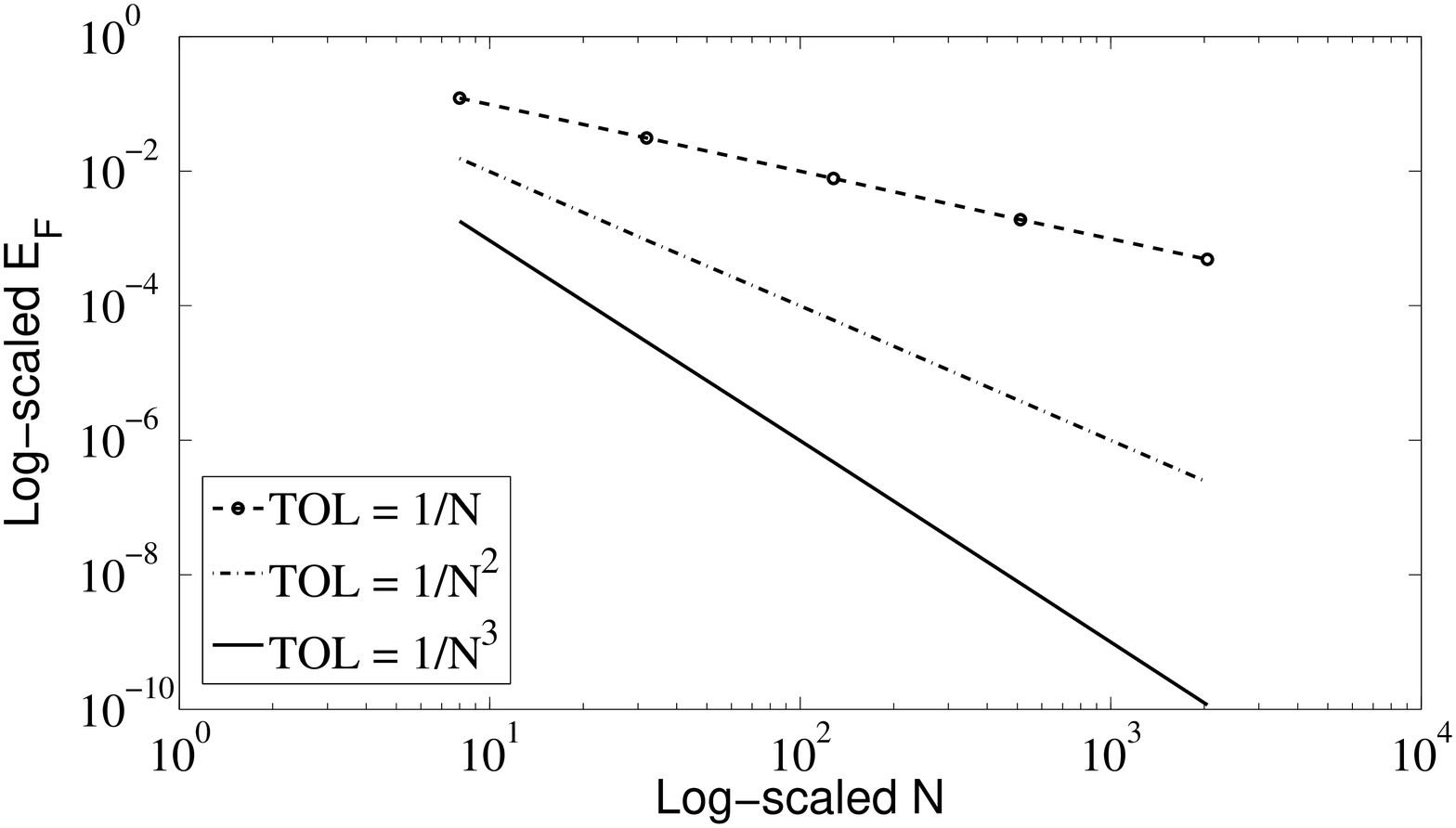}\qquad
\includegraphics[height=50mm,width=70mm]{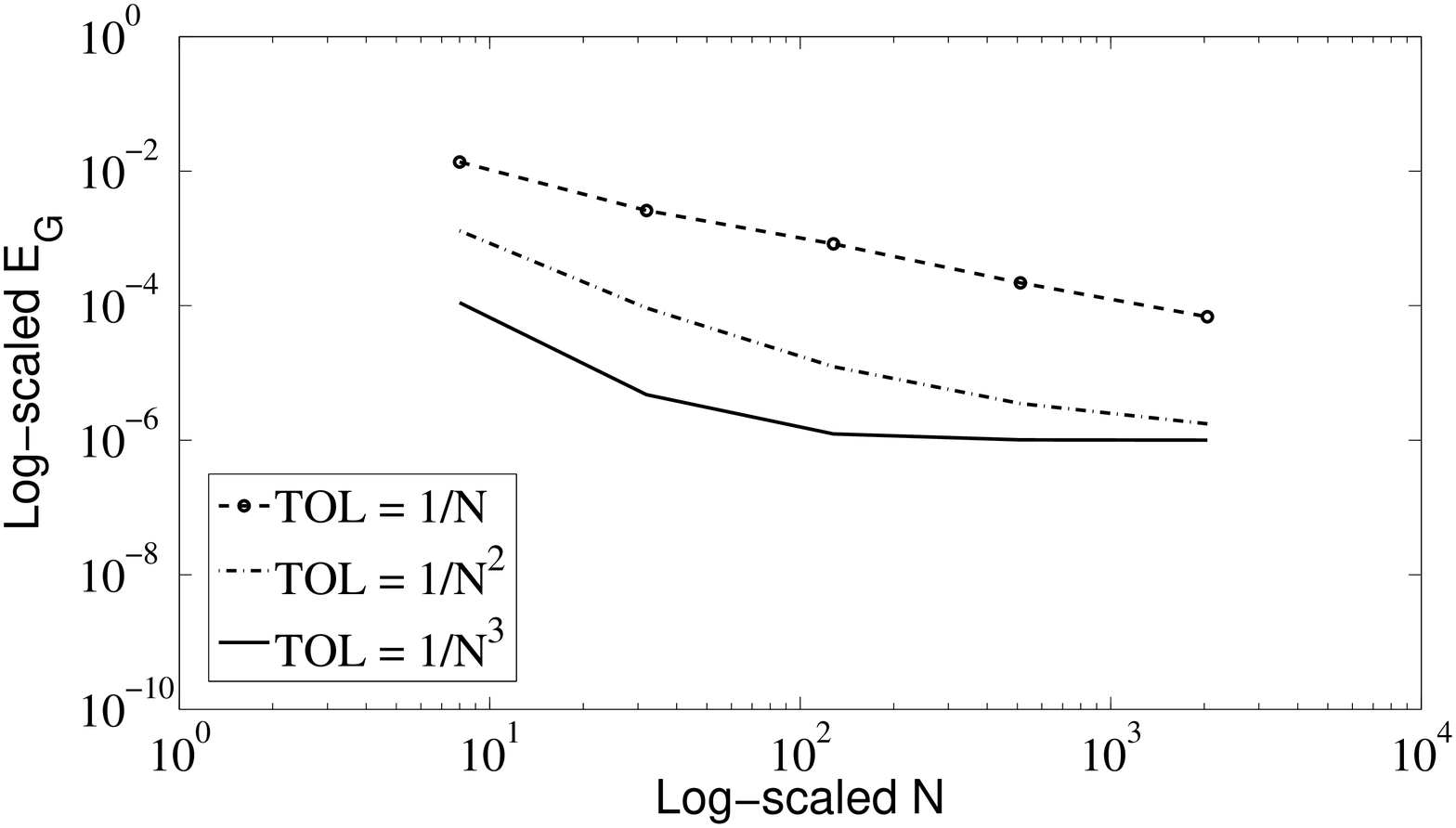}}
\caption{Error $E_F$ and $E_G$ for $P=2$, $t=1$, $0\leq
x \leq 12$ and $\beta=-\ln(10^{-6})/2T$ and different values of $TOL$. The
global error is controlled by the parameter $\beta$.}%
\label{p2alfae6}
\end{figure}

\begin{figure}[h]
\centerline{
\includegraphics[height=50mm,width=70mm]{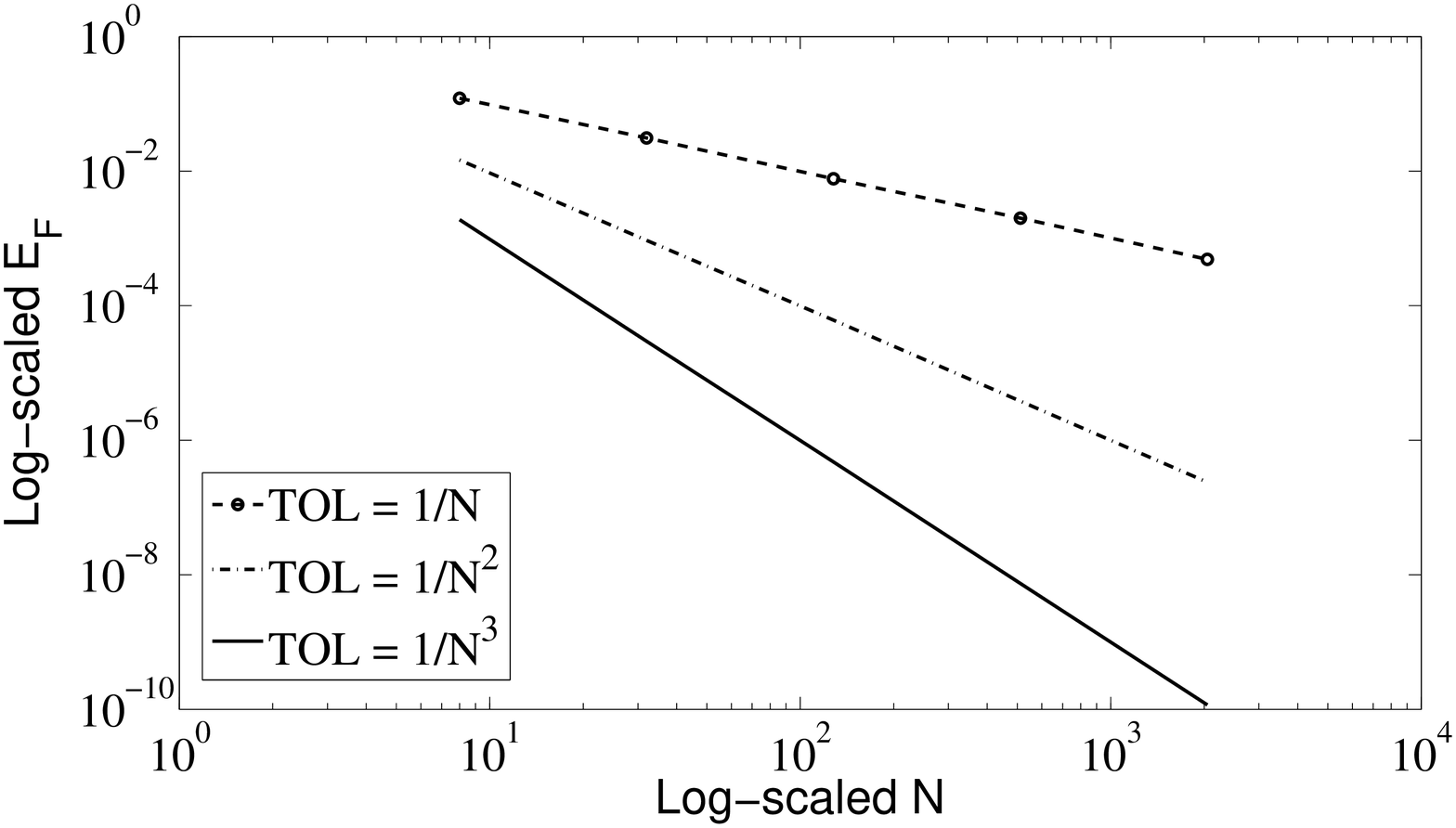}\qquad
\includegraphics[height=50mm,width=70mm]{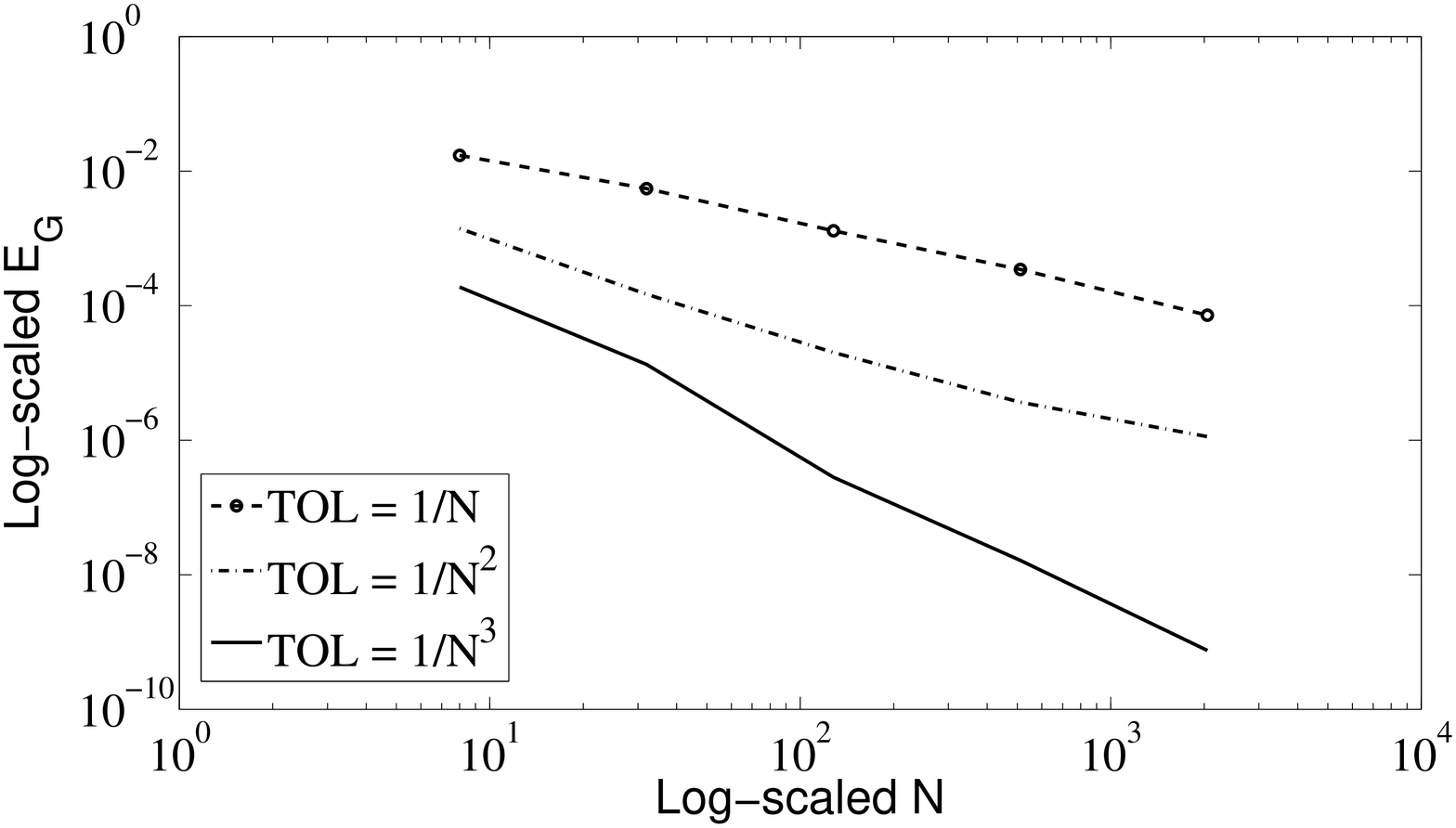}}
\caption{Error $E_F$ and $E_G$ for $P=2$, $t=1$, $0\leq
x \leq 12$ and $\beta=-\ln(10^{-10})/2T$ and different values of $TOL$. The
parameter $\beta$ is chosen such that the global error is not
affected.} \label{p2alfae10}
\end{figure}

In order to understand better how to control the trapezoidal error with
the parameter $\beta$ and how the tolerance $TOL$ affects the error, we
present  a test example which is an analytically  exactly solvable model.
We assume $P$ constant  and Fickian diffusion 
\begin{equation}
\frac{\partial n}{\partial t}
= -P\frac{\partial n}{\partial x}
+D\frac{\partial ^{2}n}{
\partial x^{2}},\mbox{ \ }x\in \left] 0, \infty \right[ ,t>0.
\label{cdex1}
\end{equation}
The initial condition is $n(x,0) = 0$,
and the boundary conditions {are}
\begin{equation}
n(0,t) = N_0, \qquad n(\infty, t) = 0.
\label{iccdex1}
\end{equation}
It will be noted that we are now considering a semi-infinite geometry.
We note the difference between this test case and our original unbound
problem. We choose this test example for two reasons: Firstly, equation
(\ref{cdex1}) can be analytically exactly solved by first applying the time-Laplace
transform and then through the inverse Laplace transform.
Secondly, this example is chosen to compare the convergence aspects of the
Laplace inversion algorithm without spatial discretization.

If we apply the Laplace transform to this problem we obtain
\begin{equation}
\widetilde{n}(x,s) = N_0 \frac{1}{s}{\rm e}^{P/2D-x\sqrt{(P/2D)^2+s}}.
\label{partsolution}
\end{equation}

The analytical solution is given by
\begin{equation}
n\left(x,t\right)=\frac{N_0}{2}\left(\mathrm{erfc}\left[\frac{x-Pt}{2\sqrt{Dt}}\right]+
{\rm
e}^{{Px}/{D}}\mathrm{erfc}\left[\frac{x+Pt}{2\sqrt{Dt}}\right]\right).
\label{AnalSolution}
\end{equation}

In Figures~\ref{p2alfae6} and ~\ref{p2alfae10}, for $P=2$, $t=1$ and
$0 \leq x \leq 12$, we plot the following errors,
\begin{equation}
E_F=\max_{1\leq i \leq N-1}| v(z,M_i)-v(z,M_{i}-1)|
\label{EF} 
\end{equation}
and
\begin{equation}
E_G = ||n(x_i,t)-\eta_{i}(t)||_\infty,
\label{EG} 
\end{equation}
where $||\cdot||_\infty$ is the
infinity norm. {We choose the interval $0 \leq x \leq 12$
  in order to avoid the influence of the right numerical boundary
  condition in the numerical computations, that in this case is $n(12,t)=0$.}

The error $E_F$ is related with the error
$E_F^M$ since we control $E_F^M$ by controlling $E_F$ with the
tolerance $TOL$.  Figures \ref{p2alfae6} and \ref{p2alfae10} show
how the parameter $\beta$, given by $\beta = - \ln(10^{-6})/2T$ in
Figure \ref{p2alfae6} and $\beta = - \ln(10^{-10})/2T$ in Figure
\ref{p2alfae10}, affects the global convergence. Note that in Figure
\ref{p2alfae6} the precision does not go further than $10^{-6}$.
The global error of Figure \ref{p2alfae6} and Figure \ref{p2alfae10}
is not affected by the spatial error $E_S$ since we apply the
Laplace inversion algorithm directly in (\ref{partsolution}).

\begin{figure}[h]
\centerline{\includegraphics[height=50mm,width=70mm]{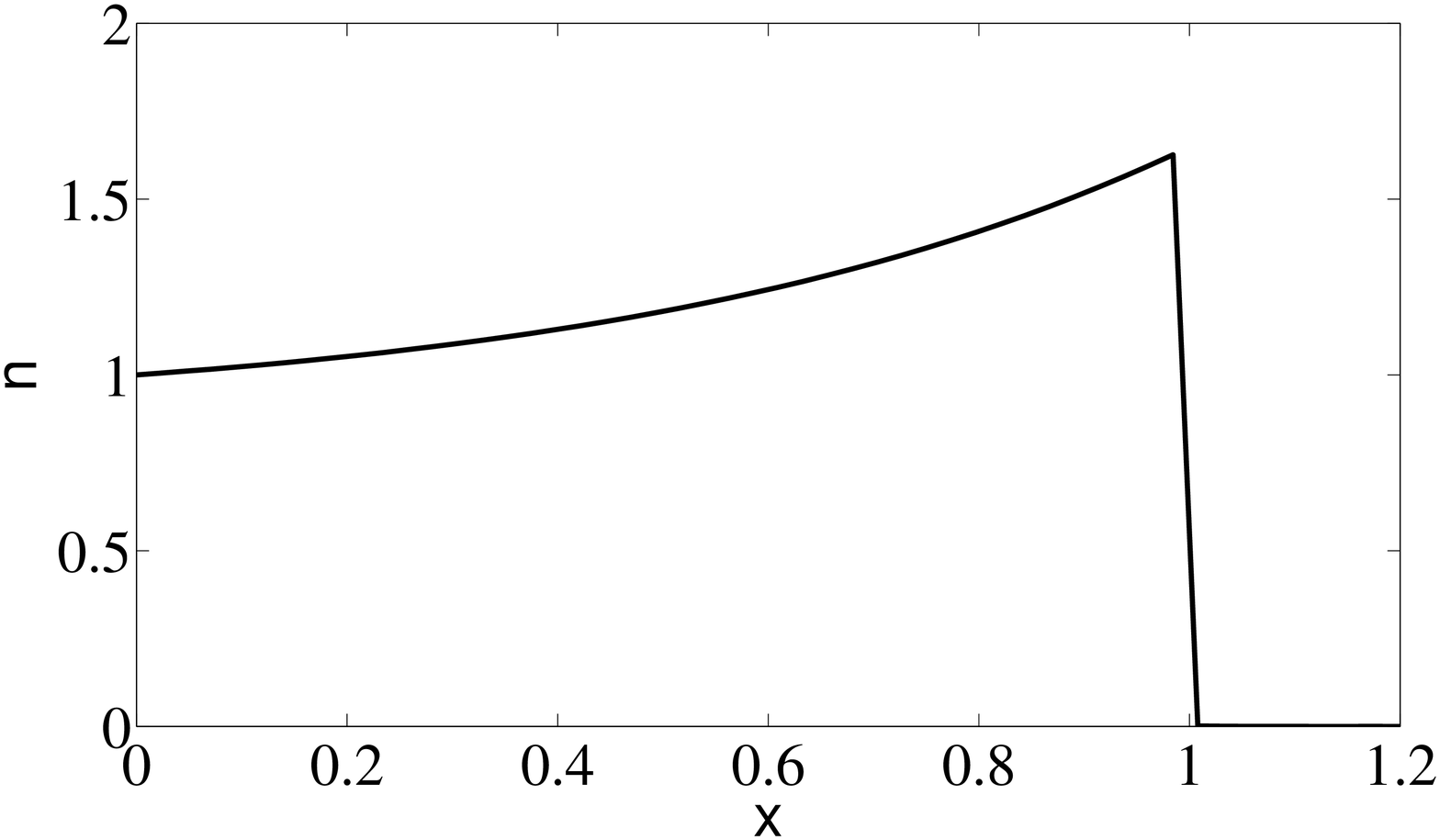}\qquad
\includegraphics[height=50mm,width=70mm]{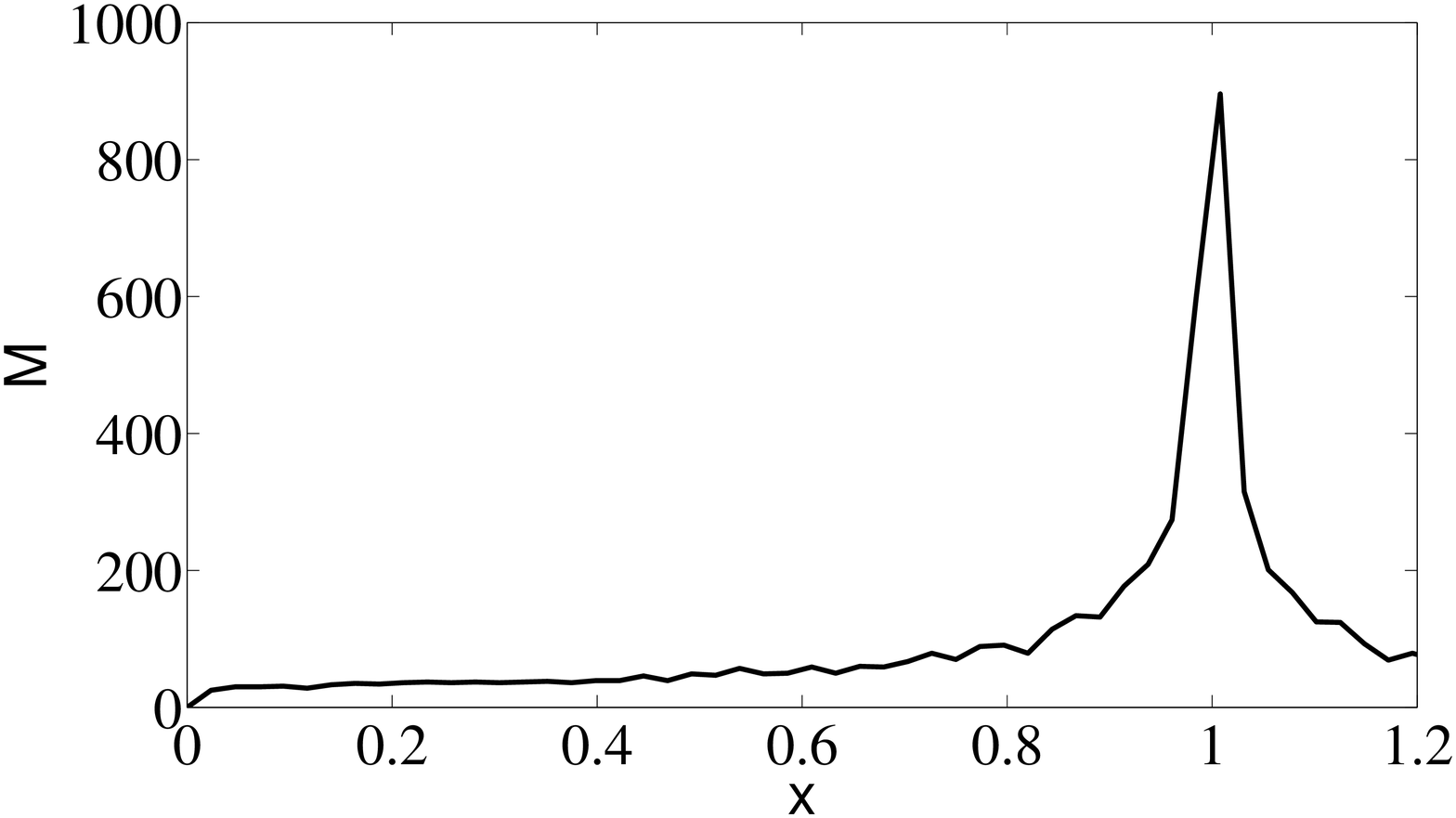}}
\caption{Number of iterations $M$ for $P=2$, $t=1$ and
$TOL=1/N^2$. Left figure: Approximate solution; Right figure: Number of iterations for each $x_i$.}
\label{iteracoes}
\end{figure}

The Laplace inversion algorithm approximates the value of the
infinite series using a truncated continued fraction and this
truncation is done by choosing an $M_i$ for each $x_i$. This $M_i$
is chosen according to which value of the tolerance $TOL$ we
consider. We show in Figure ~\ref{iteracoes} the variations of $M_i$
and it is clear the algorithm concentrates the high values of $M$ in
the region that presents steep gradients.

\

{\it Spatial discretization error  $\widetilde{E}_S(x_i,s)$}

\

{
We now turn to the discretization error $\widetilde{E}_S(x_i,s)$, defined in 
(\ref{errors}) (our main problem),
and prove that the method has truncation error of second order.}
Let us denote the differential operator $L$ given by
$$
L\widetilde{n} =\frac{d^{2}\widetilde{n}}{dx^{2}}-\lambda_s\widetilde{n}-\frac{%
d }{dx}\left( P\widetilde{n}\right).
$$
We also denote by $L_\pi$, the operator associated with the spatial discretization, given by
$$
L_\pi\widetilde{n}(x_i,s) =
\frac{\widetilde{n}_{i-1}(s)-2\widetilde{n}_i(s)+\widetilde{n}_{i+1}(s)}{\Delta
x^2}-\lambda_s\widetilde{n}_i(s)-\frac{P_{i+1}\widetilde{n}_{i+1}(s)-P_{i-1}\widetilde{n}_{i-1}(s)}{2\Delta
x},
$$
where $\widetilde{n}_i(s)$ denotes the exact solution at $\widetilde{n}(x_i,s)$.
The local truncation error is given by
$$
T_e(x_i,s) = L_\pi\widetilde{n}(x_i,s) - L\widetilde{n}(x_i,s).
$$
For a fixed $s$, we make a Taylor expansions of the functions $\widetilde{n}$ and $P$ around the point $x_i$.
We obtain, for 
a sufficiently smooth $\widetilde{n}$, 
\begin{eqnarray*}
&& K_{i,i-1}(s)\widetilde{n}_{i-1}(s)+K_{i,i}(s)\widetilde{n}_{i}(s)+K_{i,i+1}(s)\widetilde{n}_{i+1}(s)
+ (1+s)n(x_{i},0)\\
 & = &
\frac{d^2\widetilde{n}_{i}}{dx^2}(s)-\lambda_s\widetilde{n}_{i}(s)-\frac{d}{dx}(P\widetilde{n})_{i}(s)+(1+s)n(x_{i},0)\\
&&
+\left[-\frac{1}{6}P'''_i\widetilde{n}_{i}(s)-\frac{1}{2}P''_i\frac{d\widetilde{n}_{i}}{dx}(s)-\frac{1}{2}P'_i\frac{d^2\widetilde{n}_{i}}{dx^2}(s)
-\frac{1}{6}P_i\frac{d^3\widetilde{n}_{i}}{dx^3}(s)+\frac{1}{12}\frac{d^4\widetilde{n}_{i}}{dx^4}(s)\right]\Delta x^2 \\
&&+ \mathcal{O}(\Delta x^3),
\end{eqnarray*}
where $P', P'', P'''$ denotes the derivatives of $P$ (unlike in the previous test example $P$ is now not a constant).
From this result we can conclude that, for $\widetilde{n}(.,s) \in C^{4}(\mathbb{R})$, we have
$$
||T_{e}||_{\infty}=\max_{2\leq i \leq N}|T_e(x_i,s)| \leq c \Delta x^2.
$$

\begin{figure}[h]
\centerline{\includegraphics[height=55mm,width=65mm]{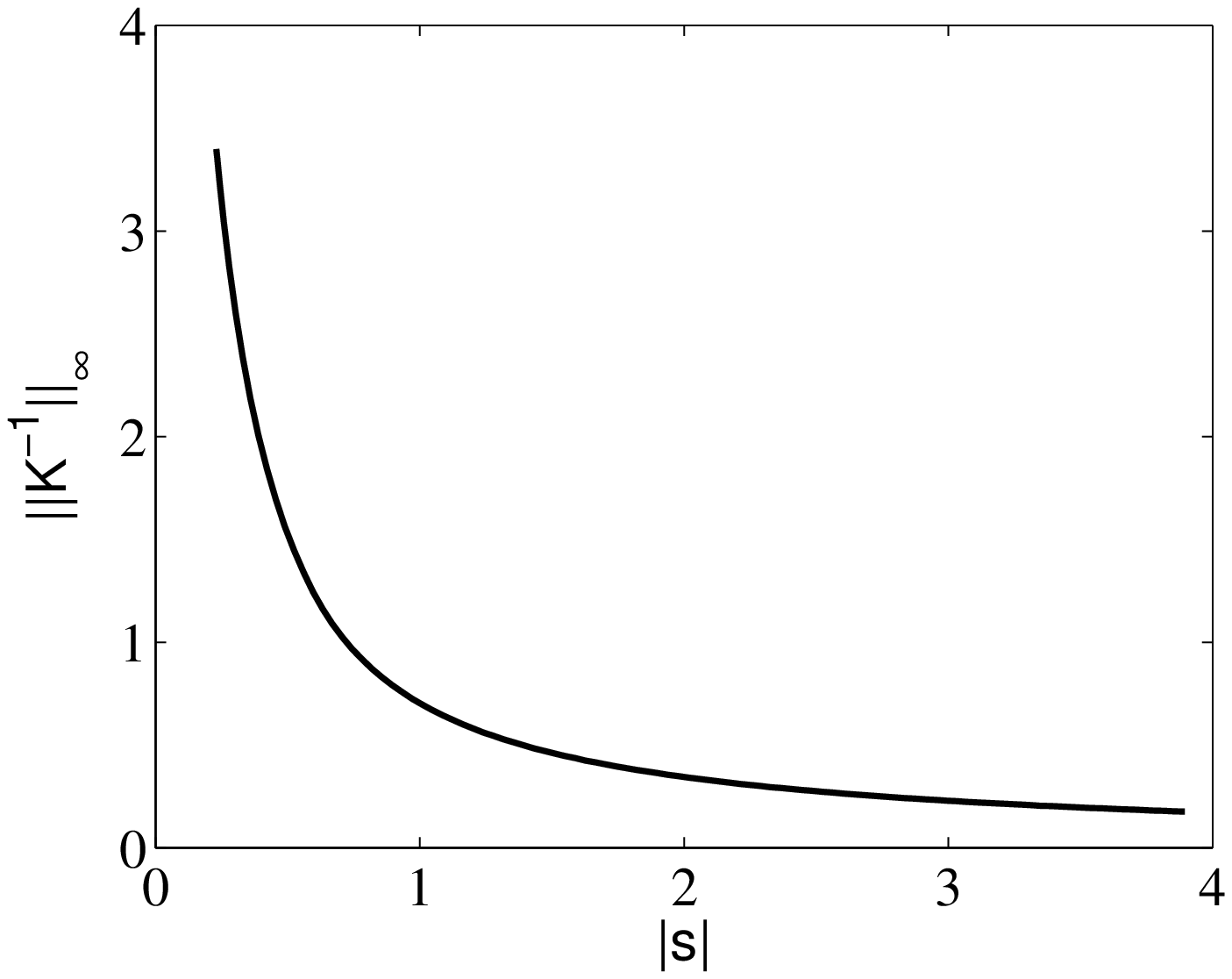}
\includegraphics[height=55mm,width=65mm]{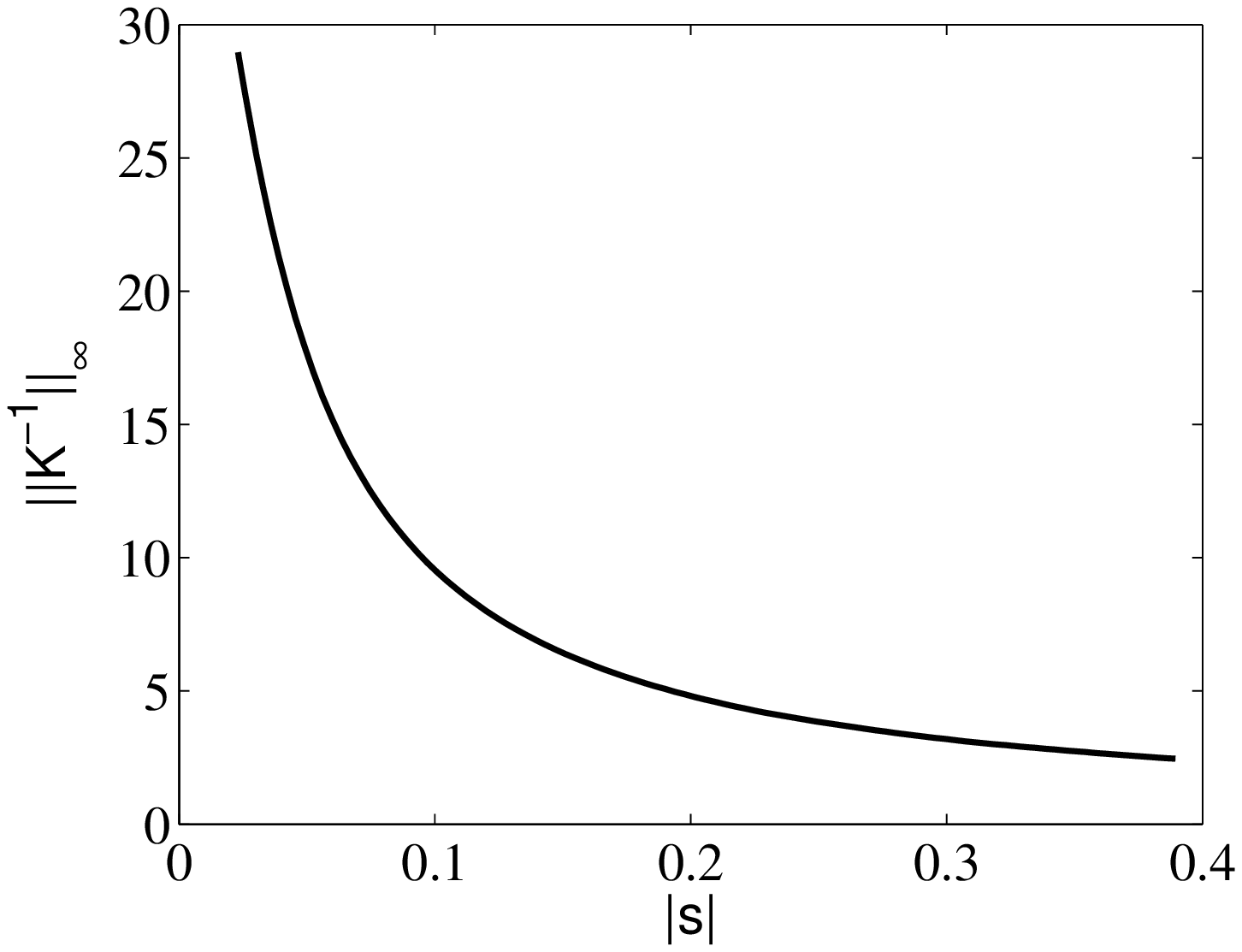}}
\caption{ Infinity norm of the matrix $K^{-1}(s)$ for $N=1000$. We have considered
$P(x)$ with $\alpha=1$.
Left figure: $T=80$; Right figure: $T=800$.}
\label{normkinv}
\end{figure}

By denoting $\widetilde{E}_i=\widetilde{E}_S(x_i,s), \ i=1, \dots, N-1$ we have
$$
L_\pi \widetilde{E}_i =T_e(x_i,s),
$$
that is,
$$
K(s)\widetilde{E}(s)=T_e(s).
$$
If $||K^{-1}(s)||_{\infty} \leq C$ then $|\widetilde{E}_i| \leq C
||T_{e}||_{\infty}$.  Since the matrix $K(s)$ is not an M-matrix
\cite{wan2007, var2000}, it is not easy to prove analytically the
inverse of $K(s)$ is bounded.  This difficulty is related to the set
of values of the parameter $\lambda_s$, given by
$$
\lambda_s = s^2+s = \beta^2 +\beta -\omega^{2} +\bold i\omega(2\beta +1), \qquad \omega=\frac{k\pi}{T}, \quad k=1, \dots, r,
$$
where $r$ defines the set of values in the Laplace domain, since for $\omega^{2} > \beta^2 +\beta$ the complex $\lambda_s$ has negative real part.
However, it is easy to see numerically that for a fixed $T$, where $T$
defines the stepsize of the trapezoidal rule used to approximate the
integral (\ref{lint}), as we refine the space step, the value
$||K^{-1}(s)||_{\infty}$ does not change significantly.  
We also observe that $||K^{-1}(s)||_{\infty}$ is larger for values of $|s|$ close to zero, indicating that
the convergence can be slower for these values, as can be observed in Figure \ref{normkinv}.

Additionally we observe that we have a similar phenomenom to the
so-called pollution effect \cite{bao2004} observed for the Helmholtz equation
and high wavenumbers where the discretization space step has to be
sufficiently refined to avoid numerical dispersion.  Also in this
context it is observed that if we have a complex number as a
coefficient in the equation, which is our case with $\lambda_s$, the
imaginary part acts as an absorption parameter, which seems to allow us
to control better the solution, decreasing the solution magnitude
\cite{fibich2003}.
Following what is reported in literature \cite{bao2004,oliveira1998}, a natural rule observed for an adjustment of the space step
is to force some relation between $T$ and the $\Delta x$. In our numerical experiments in order to retain accuracy
we have considered
\begin{equation}
\frac{t}{T} \Delta x \leq \frac{2}{100}.
\label{heuristic}
\end{equation}

\

{\it Numerical tests: Order of convergence of the numerical method}

\

In order to illustrate the behaviour of the numerical method, we consider two test problems.
First, we consider the Fickian problem (\ref{cdex1})--(\ref{iccdex1}), which exact solution is given by (\ref{AnalSolution}).
In Table 1 we show the global error (\ref{EG}), for $P=2$, $t=1$ and different values of the space step.
The value of $T=20$ was considered in order to verify (\ref{heuristic}).
These results confirm that the convergence is of second order. 
We can obtain similar results for other values of $P$.

\begin{table}[h]
\begin{center}
\begin{tabular}{l l  c}\hline
$\Delta x$ & Error  & Rate \\\hline
$10/32$    & $ 0.4000\times 10^{-2} $ & \\
$10/128$    & $ 0.2596\times 10^{-3} $ & $2.0$\\
$10/512$    & $ 0.1625\times 10^{-4} $ & $2.0$\\
$10/2048$    & $ 0.1016\times 10^{-5} $ & $2.0$\\\hline
\end{tabular}
\caption{Fickian case: Global error (\ref{EG}) for $P=2$, $t=1$, $0\leq x \leq 10$,
  $TOL=1/N^3$, $\beta=-\ln(10^{-16})/2T$, $T=20$.}
\end{center}
\label{f}
\end{table}

We now consider  a non-Fickian problem given by the telegraph equation,
\begin{equation}
\frac{\partial n}{\partial t} + \frac{\partial ^{2}n}{\partial t^{2}}
= D\frac{\partial ^{2}n}{
\partial x^{2}}, \mbox{ \ }x\in \left] 0, 2\pi \right[ ,t>0,
\end{equation}
with initial conditions
\begin{equation}
n(x,0) = \sin(\frac{x}{2}), \qquad \frac{\partial n}{\partial t}(x,0) = -\frac{1}{2}\sin(\frac{x}{2}),
\end{equation}
and  boundary conditions
\begin{equation}
n(0,t) = 0, \qquad n(2\pi, t) = 0.
\end{equation}
We can easily obtain the analytical solution given by
\begin{equation}
n\left(x,t\right)=e^{-\frac{t}{2}}\sin(\frac{x}{2}).
\label{AnalSolution2}
\end{equation}
As for the Fickian case, we present in Table 2 the global error (\ref{EG})
for $t=1$ and different space steps. We use the same value of $T$ in order to have (\ref{heuristic}).

\begin{table}[h]
\begin{center}
\begin{tabular}{l l  c}\hline
$\Delta x$ & Error & Rate \\\hline
$2\pi/32$    & $ 0.5733\times 10^{-4} $ & \\
$2\pi/128$    & $ 0.4060\times 10^{-5} $ & $1.9$\\
$2\pi/512$    & $ 0.2397\times 10^{-6} $ & $2.0$\\
$2\pi/2048$    & $ 0.1627\times 10^{-7} $ & $1.9$\\\hline
\end{tabular}
\caption{Non-Fickian case: Global error (\ref{EG}) for $t=1$, $0\leq x \leq 2\pi$,
  $TOL=1/N^3$, $\beta=-\ln(10^{-16})/2T$, $T=20$.}
\end{center}
\label{nf}
\end{table}

It will be noted that our main problem is unbounded. But with at least one
zero boundary condition for each, the two test examples ( Fickian and
non-Fickian), although semi-bounded and bounded respectively, can be
computationally viewed as similar to our main problem. We observe from
Tables \ref{f} and \ref{nf} for the test examples that we obtain second order convergence
as predicted by the theoretical analysis for the main problem.

\section{Numerical results for $n(x,t)$,  $j(x,t)$ and $<x^2(t)>$}

To do the numerical experiments we consider the equations
\begin{equation}
\frac{\partial^{2}n}{\partial t^{2}}+\frac{\partial n}{\partial t}
= -\frac{\partial}{\partial x}(Pn)+\frac{\partial ^{2}n}{
\partial x^{2}}
\label{equationnr}
\end{equation}
and
\begin{equation}
j+\frac{\partial j}{\partial t} = -\frac{\partial n}{\partial x}+Pn,
\label{fluxnr}
\end{equation}
for
$$
P(x)=-\frac{dV}{dx} \qquad
\mbox{and} \qquad V(x;\alpha)=\frac{1}{J_0(\bold i\alpha)}e^{\alpha\cos{x}}-1.
$$

\begin{figure}[h]
\centerline{
\includegraphics[height=55mm,width=65mm]{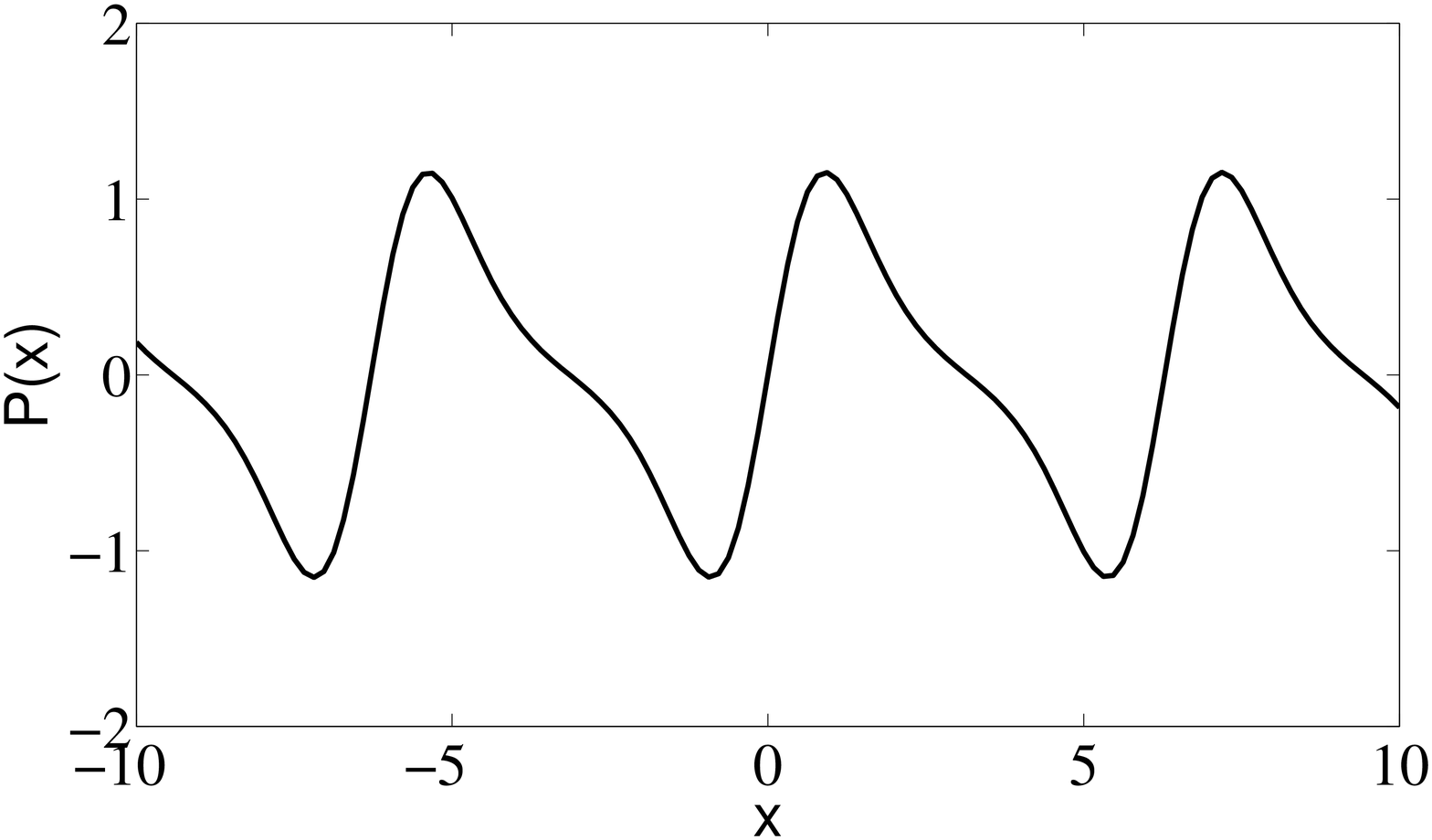}
\includegraphics[height=55mm,width=65mm]{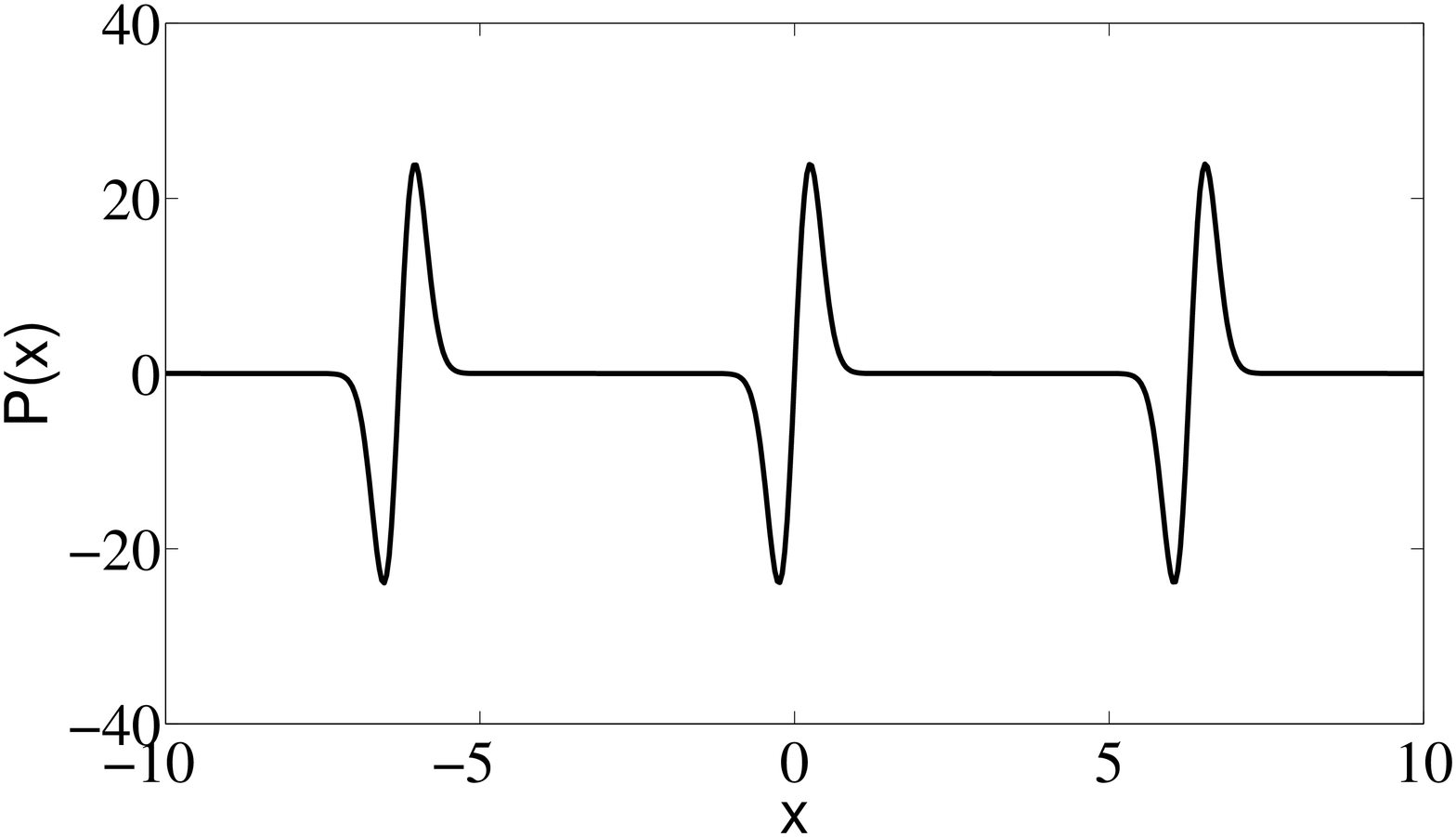}
}
\caption{Potential $P(x)$. Left figure: $\alpha=1$; Right figure: $\alpha=16$.}
\label{p1-16}
\end{figure}

For $\alpha = 1$, the potential is smoother compared with $\alpha=16$, as it
is shown in Figure \ref{p1-16}. In Figure \ref{p1-16} we observe that
$P(x)$ for $\alpha=1$ changes between $-1$ and $1$, whereas  for $\alpha=16$ changes
between $-20$ and $20$ and the change is not smooth. Our method can deal
very well with both cases.

We consider the initial conditions,
\begin{eqnarray}
n\left(x,0\right) &=&\frac{1}{L \sqrt{\pi}}e^{-x^2/L^2},{\ \ \ \ }
\frac{\partial n }{\partial t}\left( x,0\right) = 0,
\\
j\left(x,0\right) &=&\frac{1}{L
  \sqrt{\pi}}e^{-x^2/L^2}\left(\frac{1}{L^2}2x+
  P(x)\right),
\end{eqnarray}
and the boundary conditions are given by
$$
\lim_{x\rightarrow \infty} n(x,t)=0 \quad \mbox{and} \quad \lim_{x\rightarrow -\infty}n(x,t)=0.
$$
Note that the stationary solution of the problem is given by
$$
n_{st}(x)=N_r\exp\left(-V(x)\right),
$$
where $N_r$ is a normalization value.

\begin{figure}[h]
\centerline{
\includegraphics[height=55mm,width=65mm]{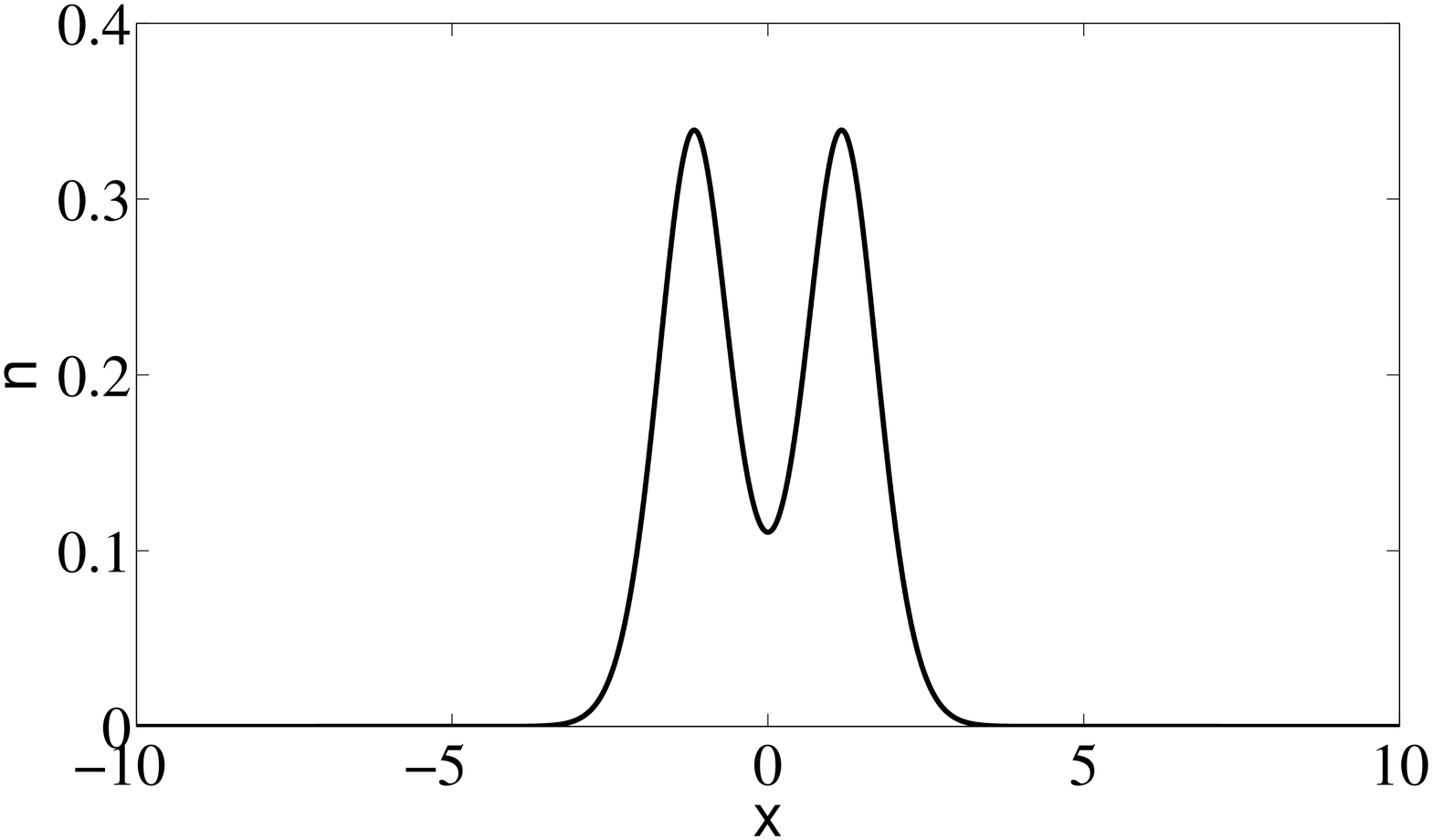}
\includegraphics[height=55mm,width=65mm]{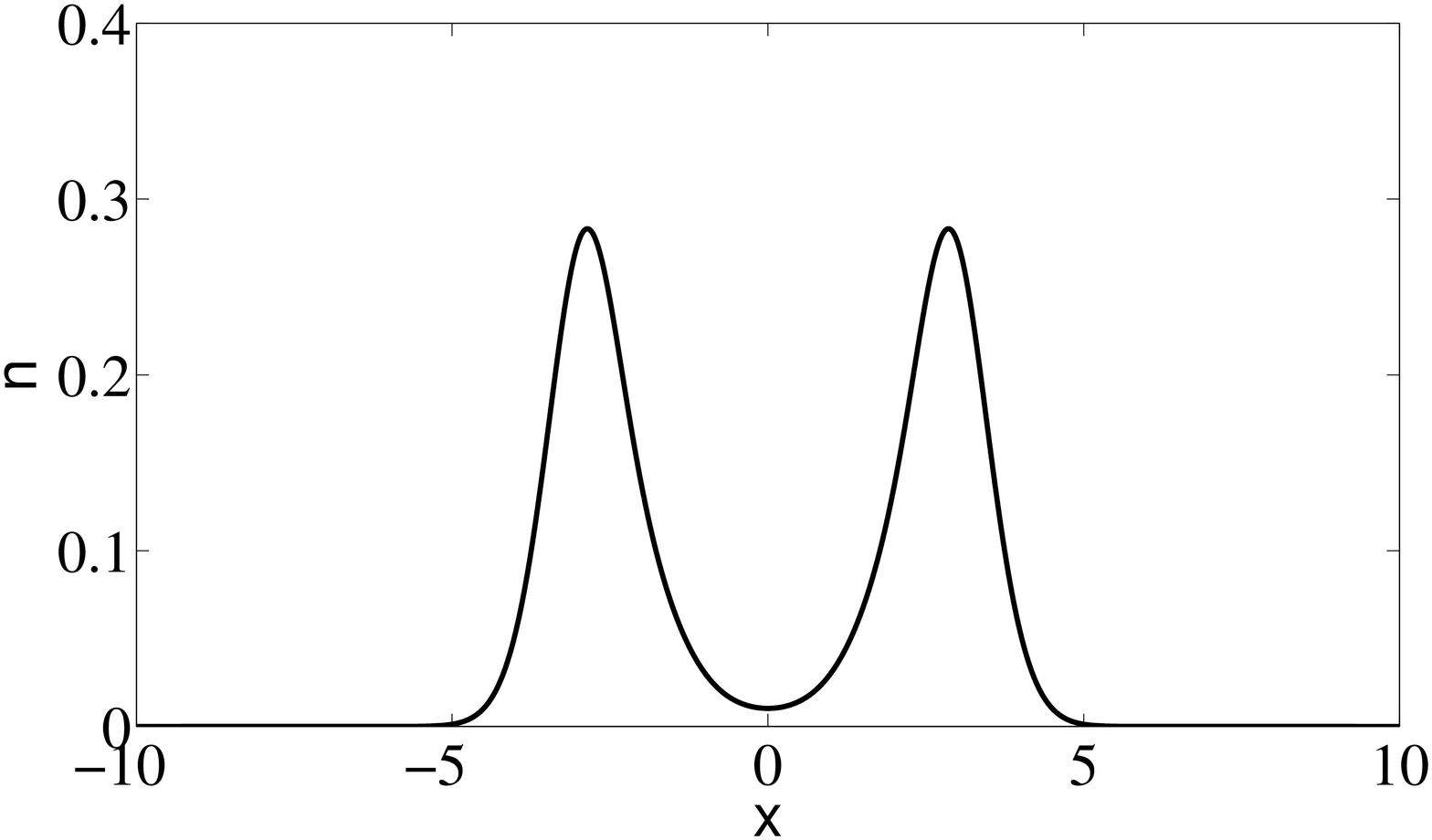}}
\caption{Particle density $n(x,t)$ for $\alpha=1$. Left figure: Curve for
  instant of time $t=1$; Right figure: Curve
  for instant of time $t=3$.}
\label{alpha1-1-3}
\end{figure}
\begin{figure}[h]
\centerline{\includegraphics[height=55mm,width=65mm]{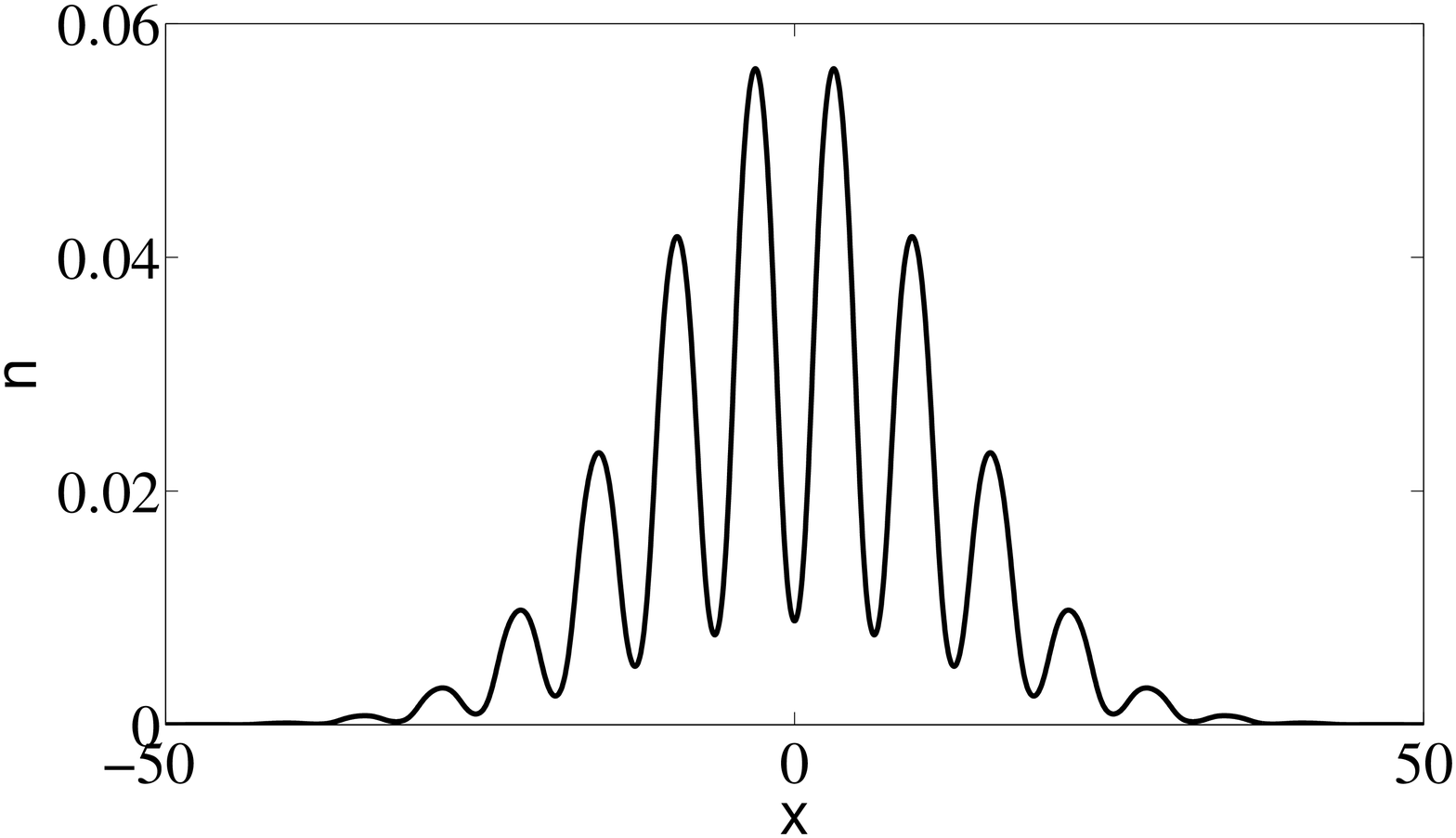}
\quad \includegraphics[height=55mm,width=65mm]{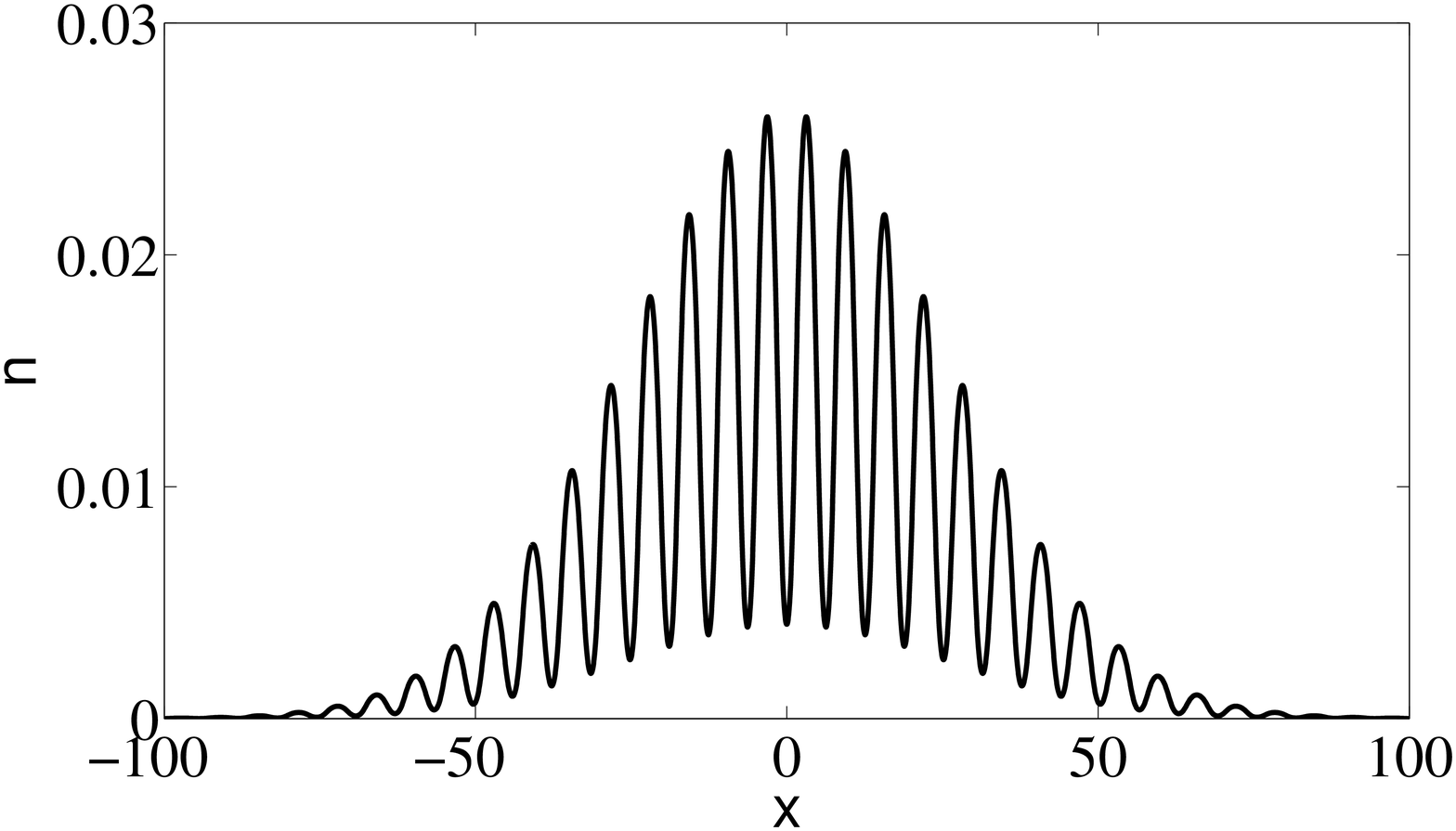}}
\caption{Particle density $n(x,t)$ for $\alpha=1$. Left figure: Curve for instant of time
$t=100$ ;Right figure: Curve for instant of time
$t=500$.}
\label{alpha1-100-500}
\end{figure}
For $\alpha=1$ we show in Figures \ref{alpha1-1-3} and \ref{alpha1-100-500}
the behaviour of the solution as we increase time from $t=1$ until $t=500$.
The peak starts to split into two and then we have
several waves forming that goes to the right and left.
The domain where the function is not zero becomes
larger as we travel in time. For that reason the computational domain
increases considerably which requires more computational effort regarding the
discretization
in space. For an iterative method where we need to consider a discretization in time, it would require
more computational effort for long times as we need to iterate in time
whereas the Laplace transform has the advantage of not
iterating in time and therefore it is the same if we compute the solution for
short times or long times.

\begin{figure}[h]
\centerline{
\includegraphics[height=55mm,width=65mm]{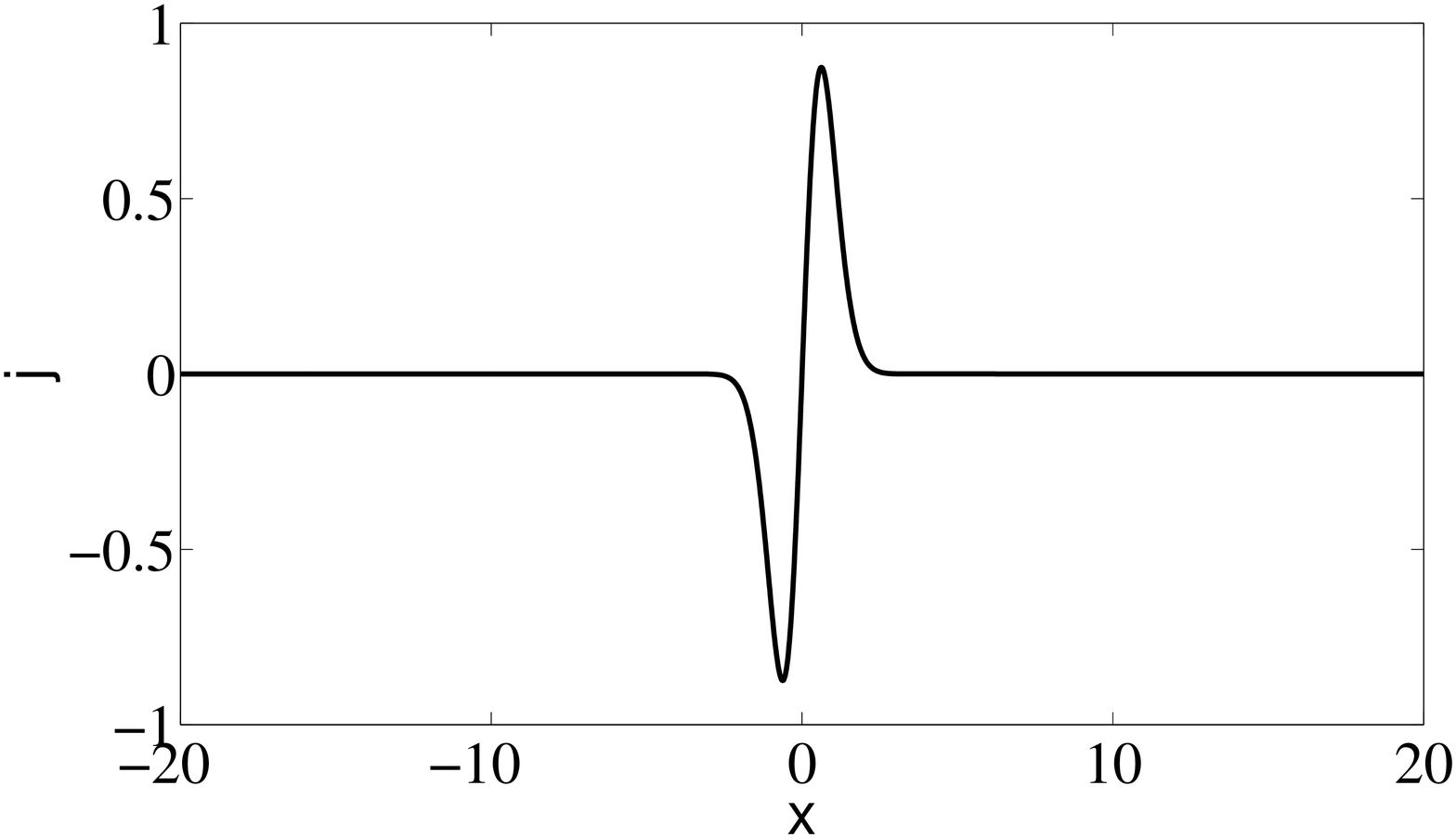}
\includegraphics[height=55mm,width=65mm]{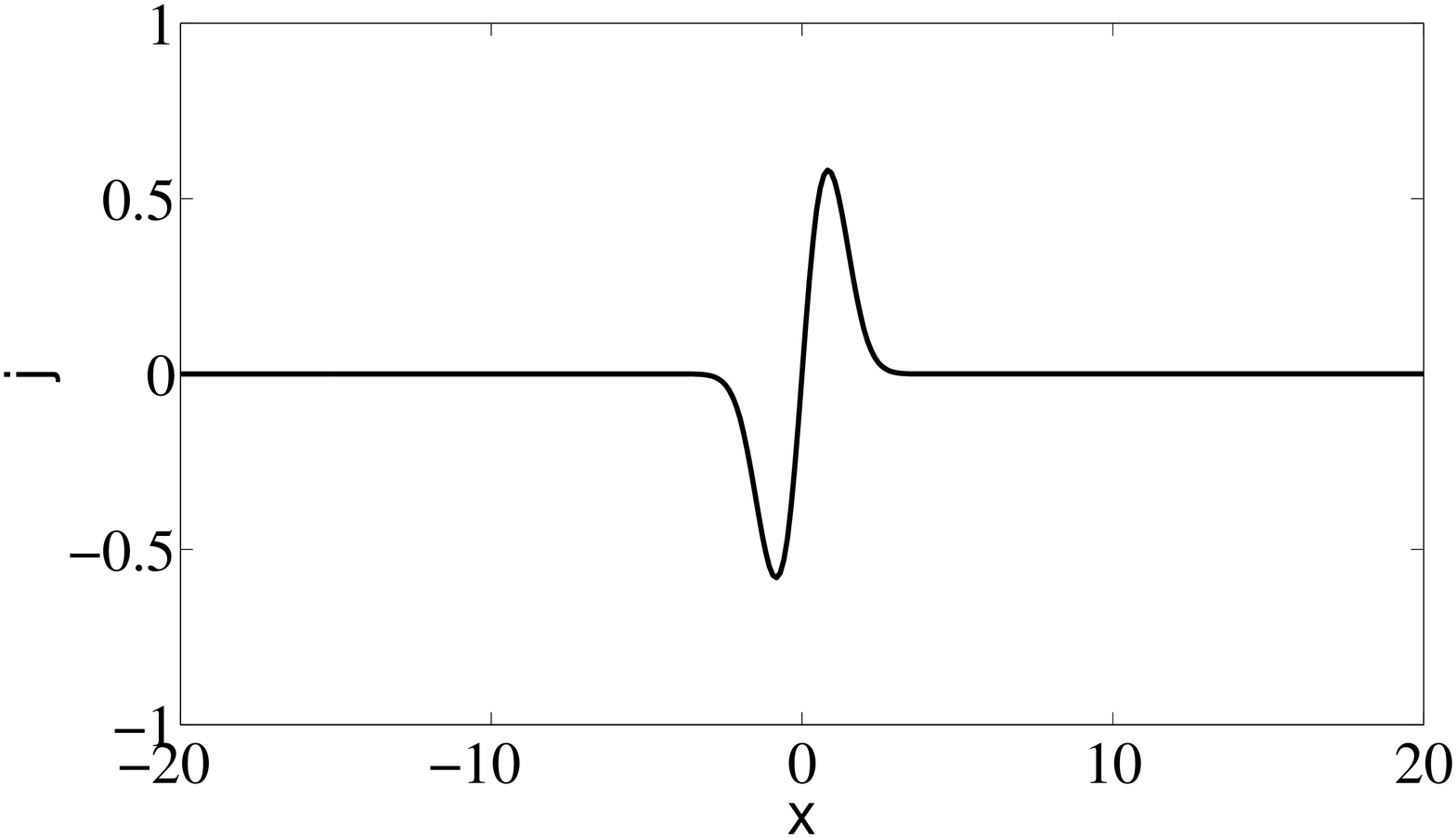}}
\caption{Density flux $j(x,t)$ for $\alpha=1$. Left figure: Curve for
  instant of time $t=0$; Right figure: Curve
  for instant of time $t=1$.}
\label{j-0-1}
\end{figure}
\begin{figure}[h]
\centerline{
\includegraphics[height=55mm,width=65mm]{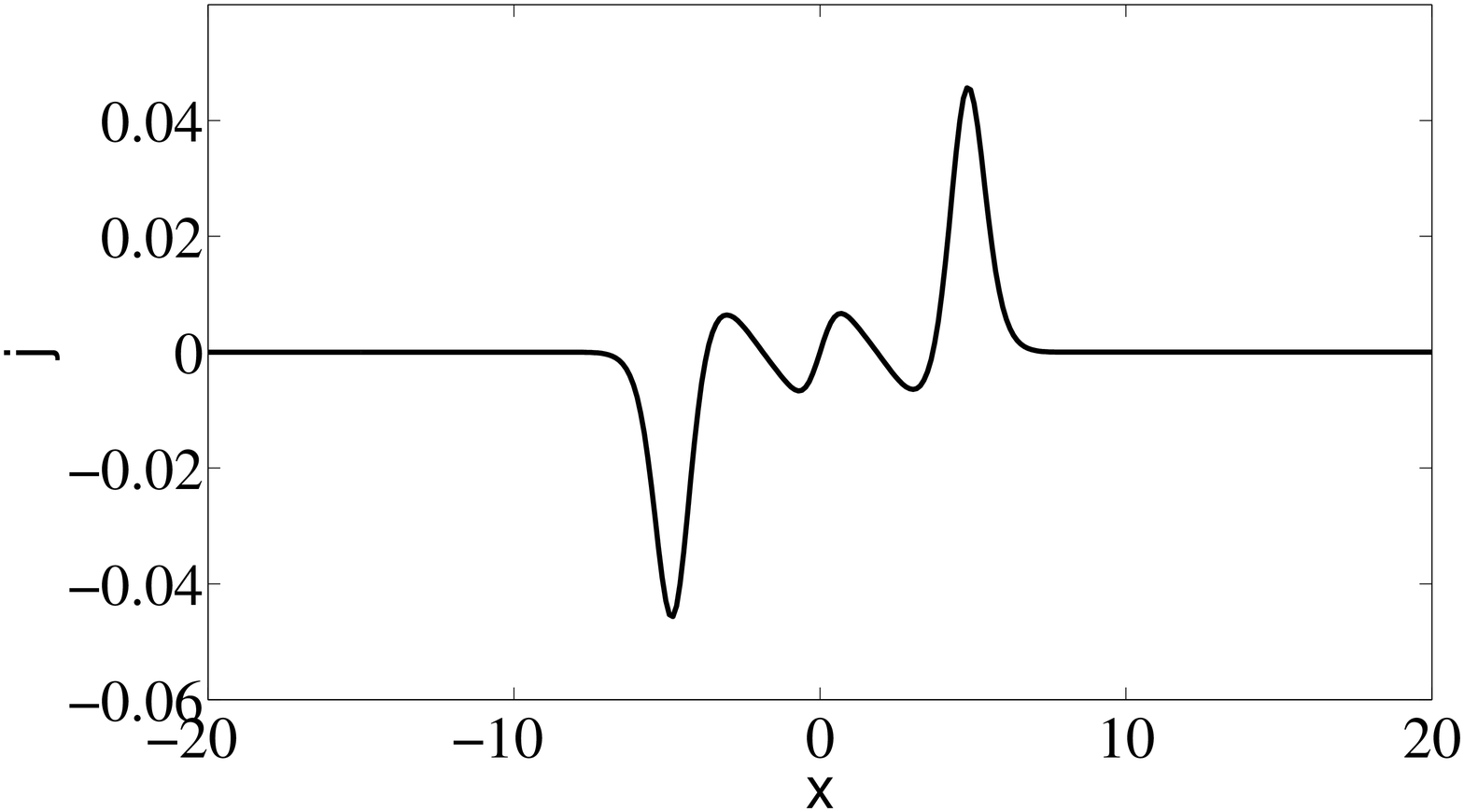}
\includegraphics[height=55mm,width=65mm]{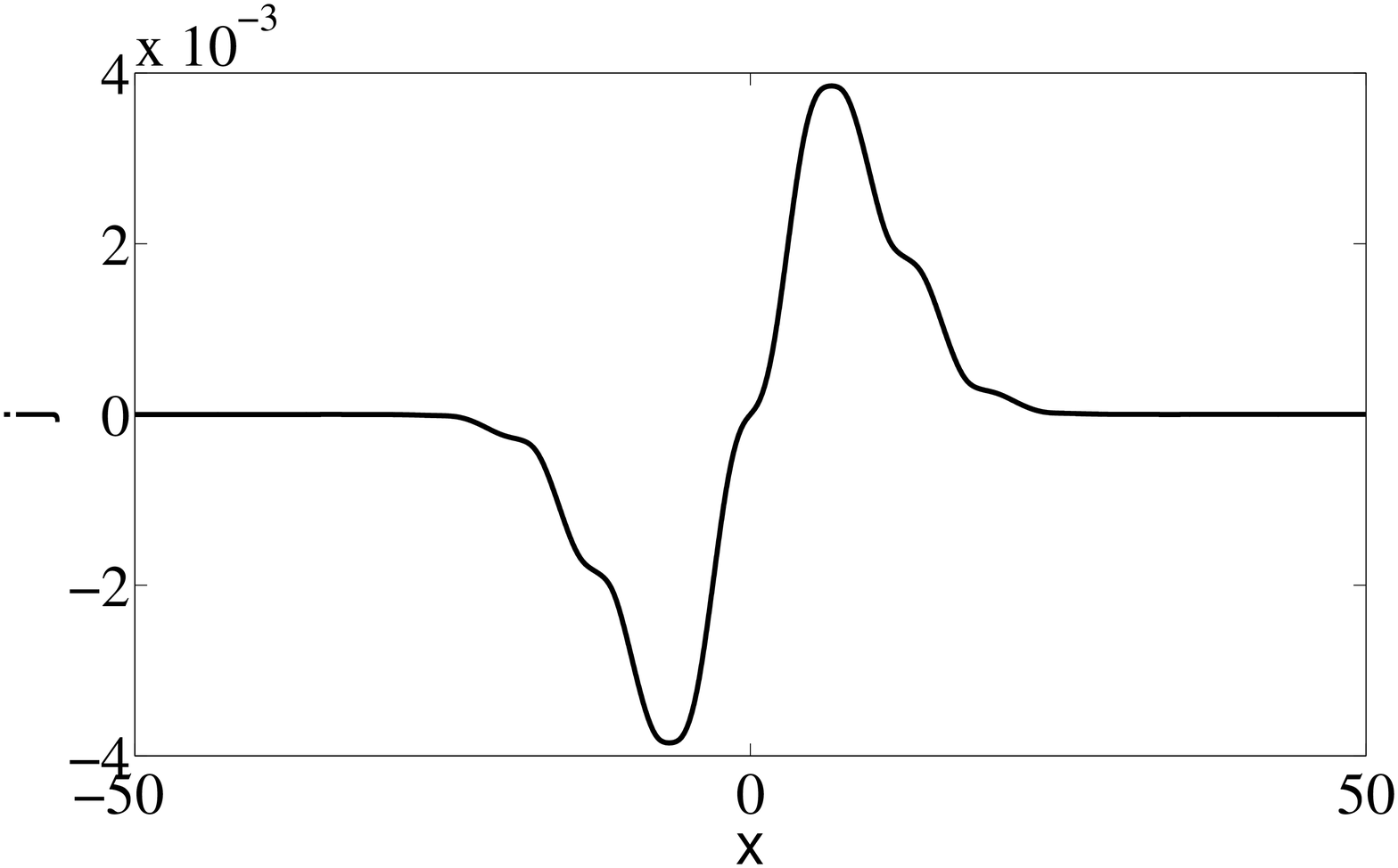}}
\caption{Density flux $j(x,t)$ for $\alpha=1$. Left figure: Curve for
  instant of time $t=5$; Right figure: Curve
  for instant of time $t=30$.}
\label{j-5-30}
\end{figure}

In Figure \ref{j-0-1} we plot the flux for $\alpha=1$, as it evolves from $t=0$ to
$t=1$ and in Figure \ref{j-5-30} as it evolves  from $t=5$ to $t=30$.


A quantity of physical interest in diffusion problems is
the mean square displacement defined by
 $$<x^2(t)> = \int_{-\infty}^{\infty}{[x^2n(x,t)]} dx.$$
 For the Fickian case, $<x^2(t)>$ is linear in $t$ for all times in
 the absence of a potential.  Now we would like to present
 calculations of $<x^2(t)>$ for the non-Fickian diffusion. At short
 times, and in the presence of a potential, the mean square
 displacement, $<x^2(t)>$, shows a $t^2$ behaviour, see Figure 11.
 This is due to inertial effects which are captured by a non-Fickian
 diffusion equation.  

\begin{figure}[h]
\centerline{\includegraphics[height=55mm,width=65mm]{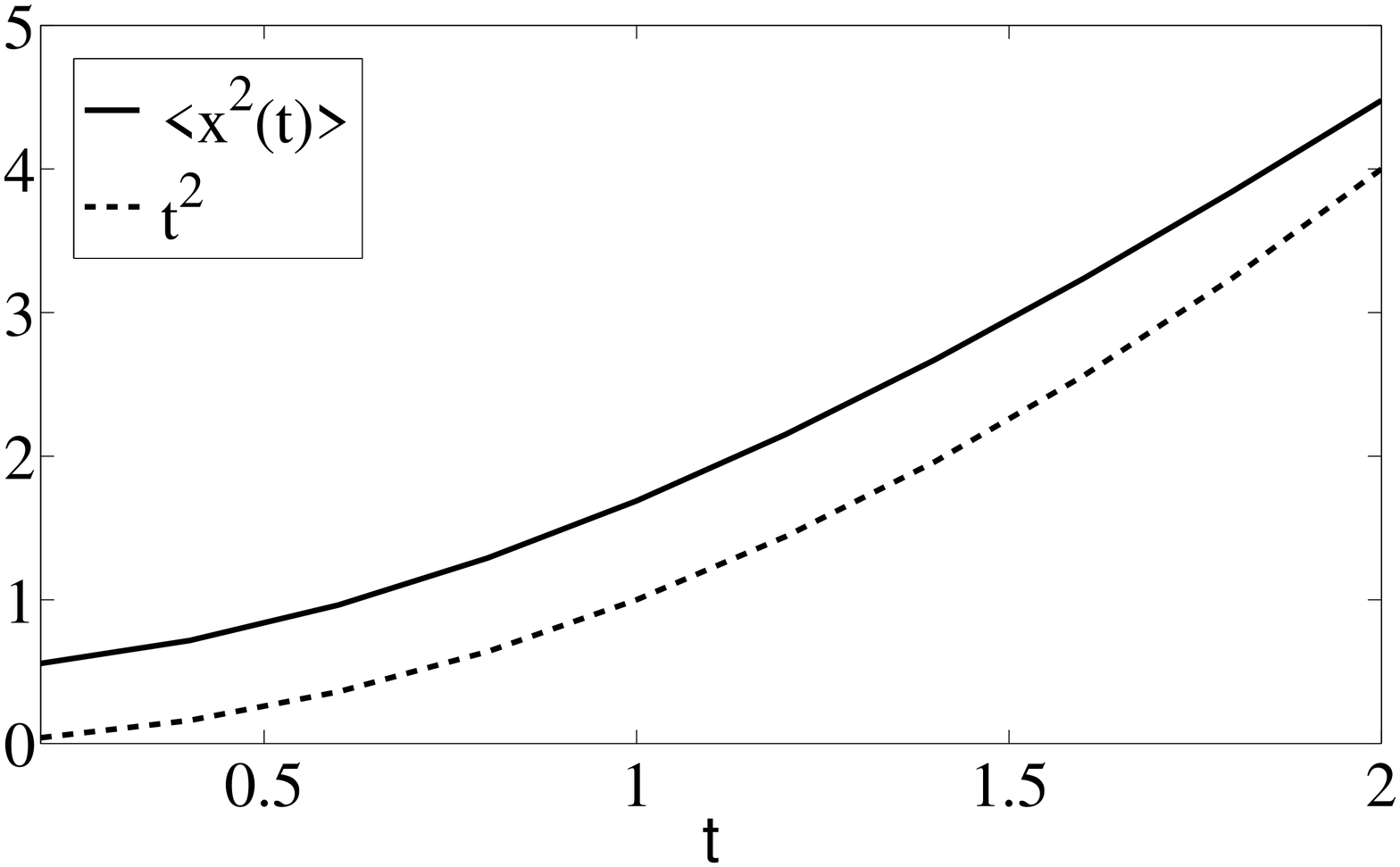}
\quad \includegraphics[height=55mm,width=65mm]{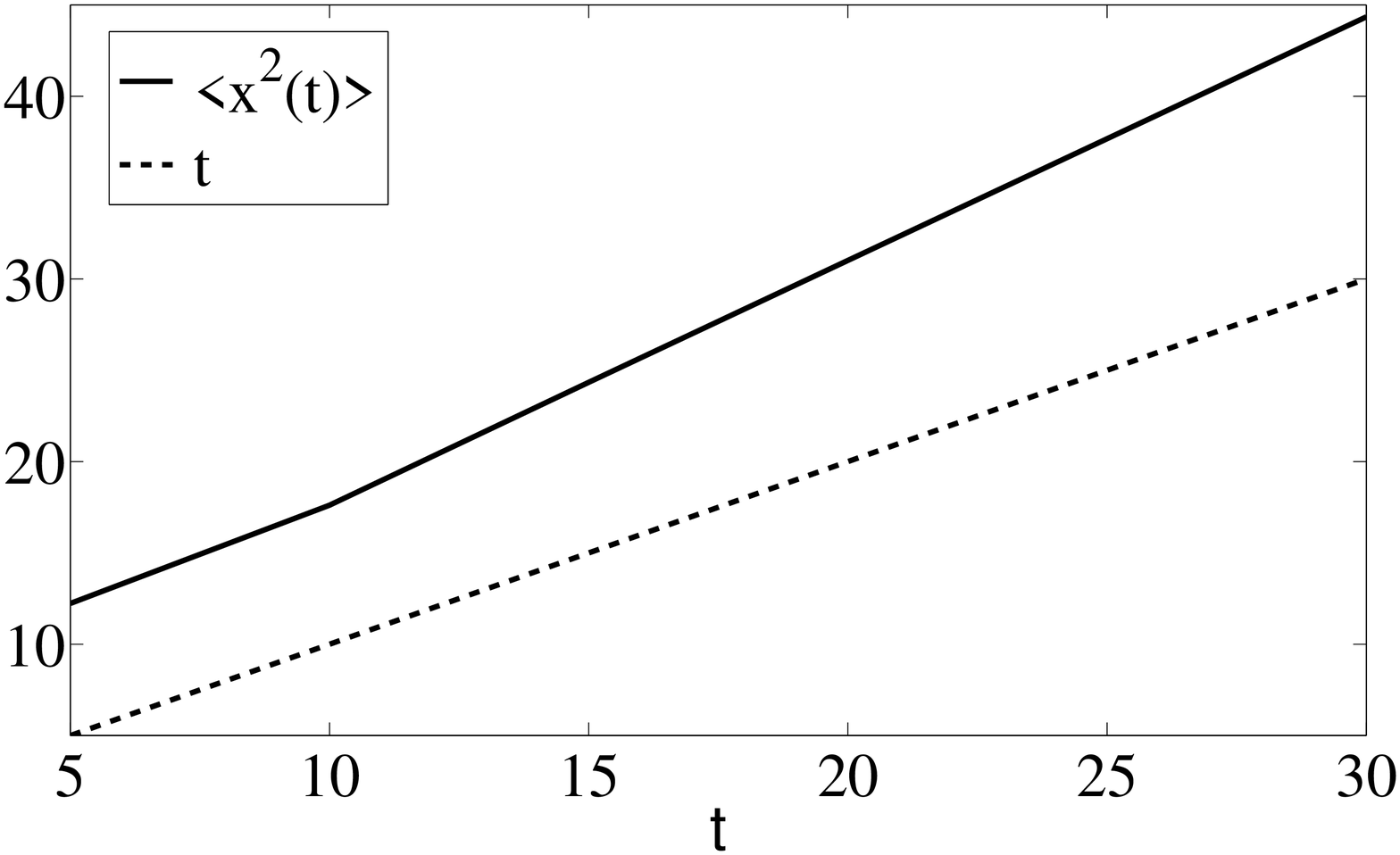}}
\caption{Mean square displacement for $\alpha=1$; Left figure:
  Curves for $t\in [0,2]$; Right
figure: Curves for $t\in[5,30]$.}
\label{alpha16-500-1000-5000}
\end{figure}

\begin{figure}[h]
\centerline{
\includegraphics[height=55mm,width=65mm]{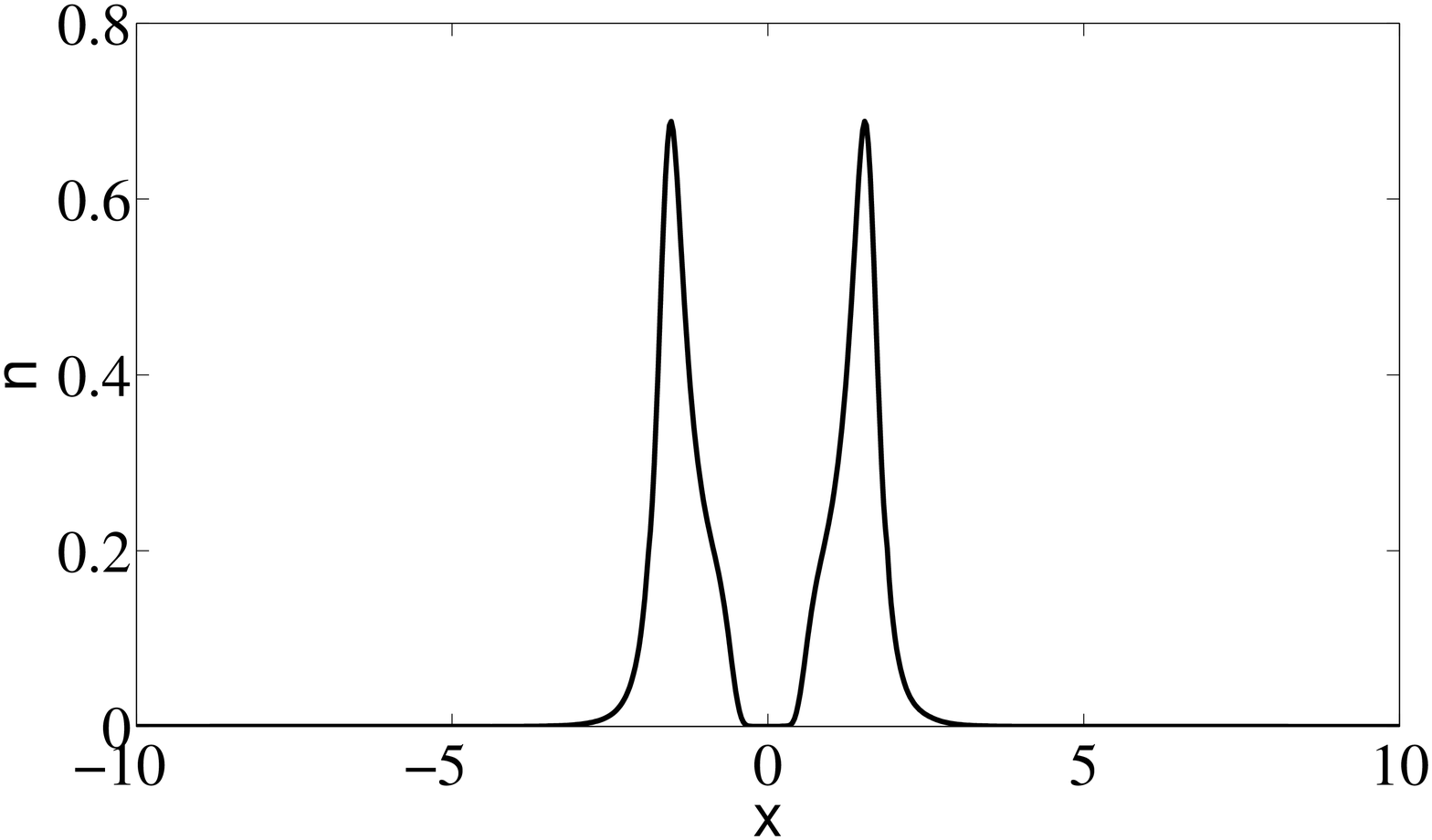}
\includegraphics[height=55mm,width=65mm]{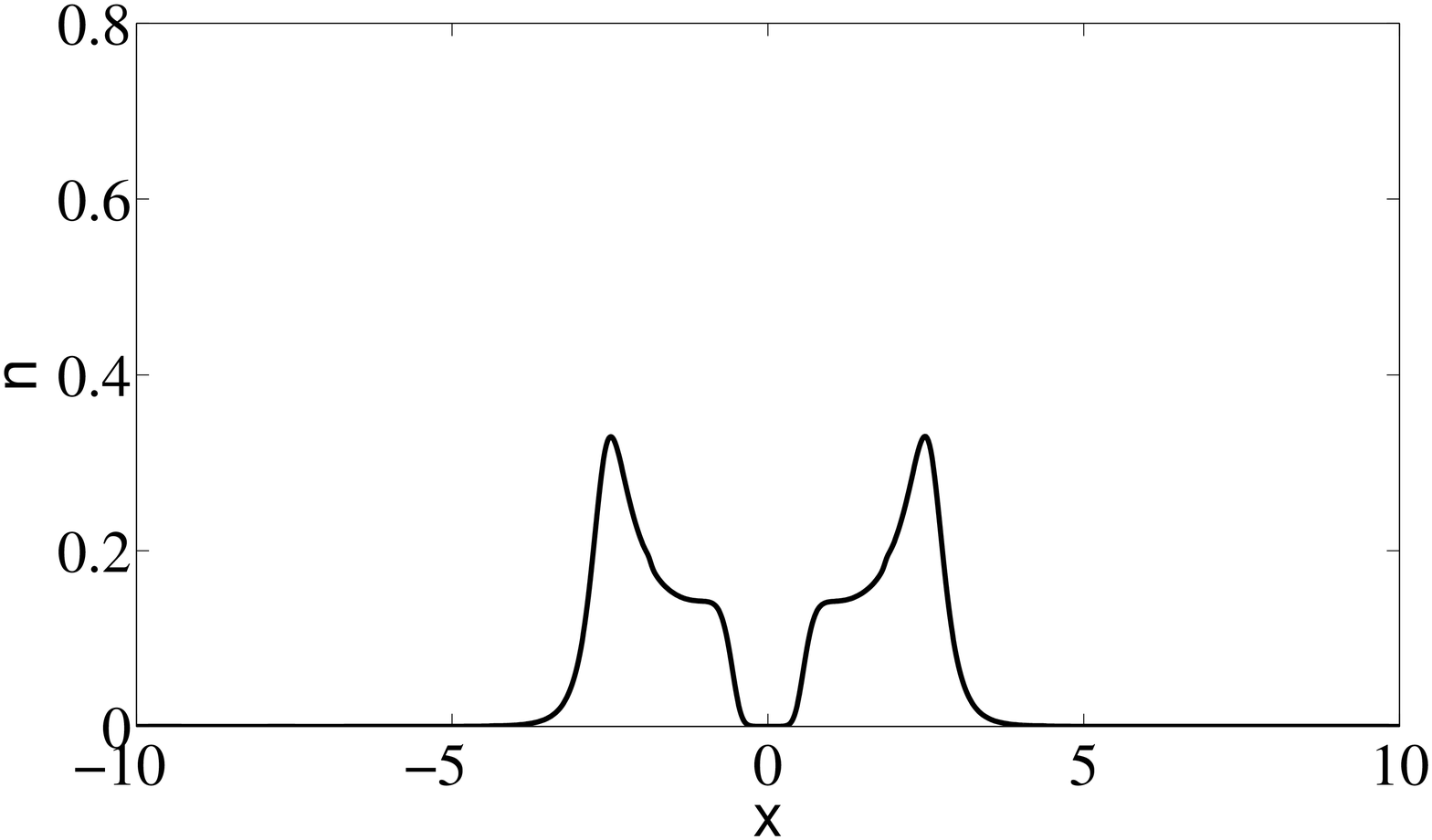}}
\caption{Particle density $n(x,t)$ for $\alpha=16$. Left figure: Curve for
  instant of time $t=1$; Right figure: Curve
  for instant of time $t=2$.}
\label{alpha16-1-2}
\end{figure}

For $\alpha = 16$ we show the evolution of the solution in the first
instants of time. We see the solution presents very steep gradients
and the method is able to give accurate solutions.
First we observe
how the wave split for $t=1$ and $t=2$ in Figure \ref{alpha16-1-2}.

Next in Figure \ref{alpha16-500-1000-5000} we observe the behaviour for very
large times. It is interesting to see how the Laplace method is able to give
very quickly solutions for very large times. 
An iterative numerical method in time,
it would
take a large amount of time to run experiments for such long times such as
$t=5000$
or $t=10000$ as we can see in Figure \ref{alpha16-10000}.
The flux for $\alpha=16$ is plotted in Figure \ref{jalpha16t1}.

\begin{figure}[h]
\centerline{\includegraphics[height=55mm,width=65mm]{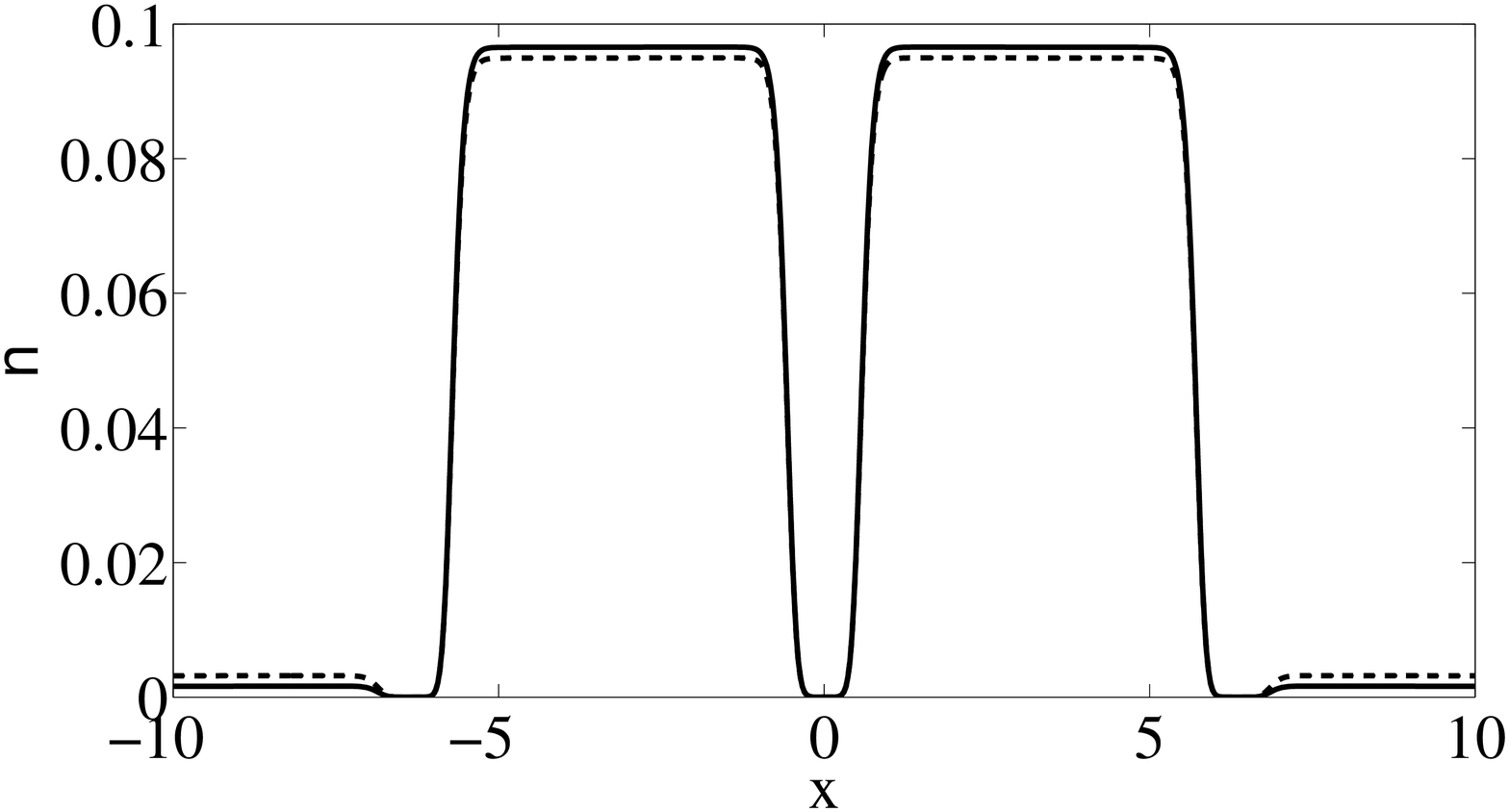}
\quad \includegraphics[height=55mm,width=65mm]{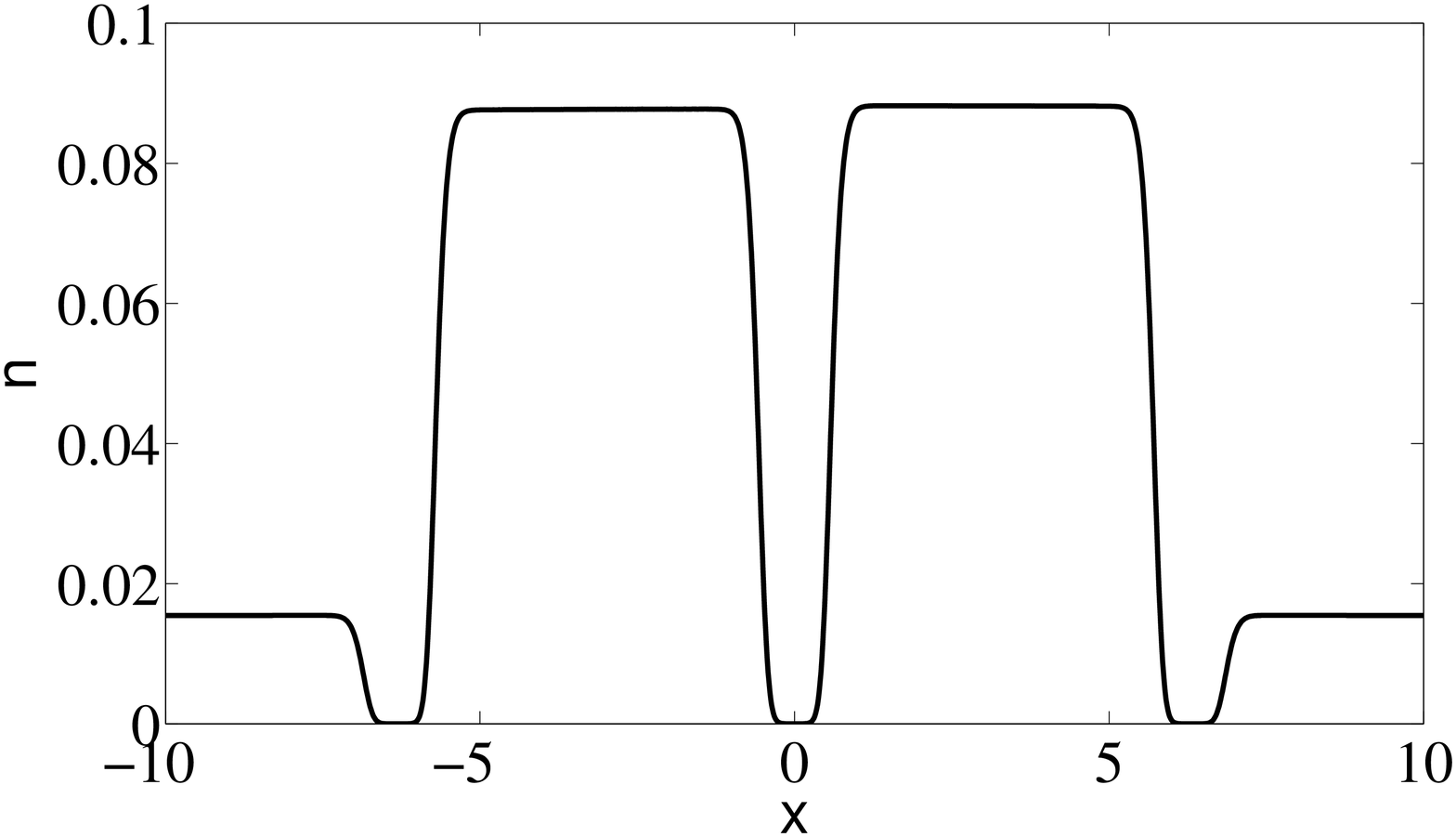}}
\caption{Particle density $n(x,t)$ for $\alpha=16$. Left figure: Curves for instant of times
$t=500$ $(\--)$, $t=1000$ $(\-- \--)$; Right figure: Curve for instant of time
$t=5000$ $(\--)$.}
\label{alpha16-500-1000-5000}
\end{figure}

\begin{figure}[h]
\centerline{ \includegraphics[height=55mm,width=65mm]{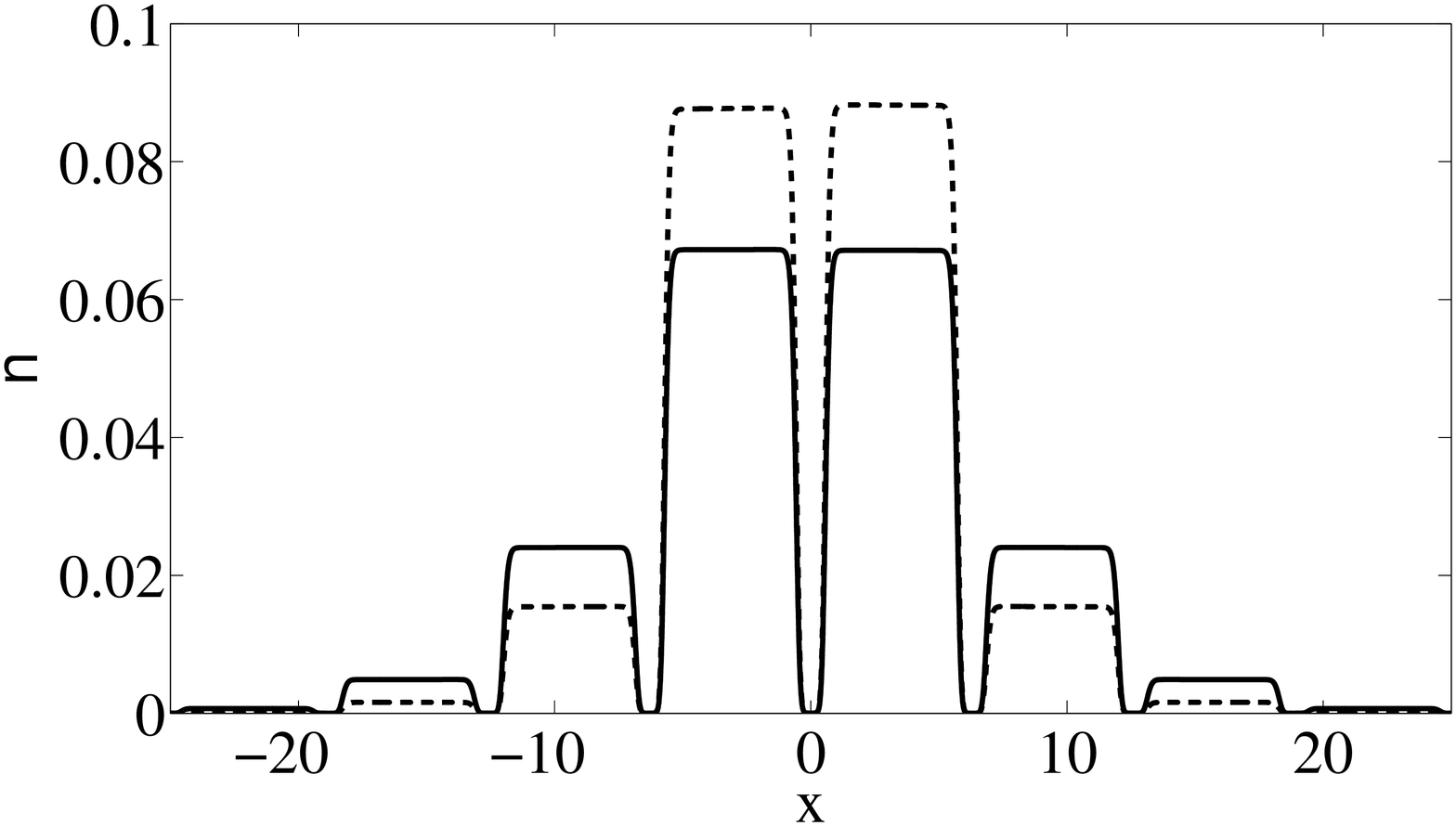}}
\caption{Particle density $n(x,t)$ for $\alpha=16$. Curves for instant of time $t=5000$ $(\-- \--)$;
$t=10000$ $(\--)$.}
\label{alpha16-10000}
\end{figure}

\begin{figure}[h]
\centerline{
\includegraphics[height=6.0cm,width=7.0cm]{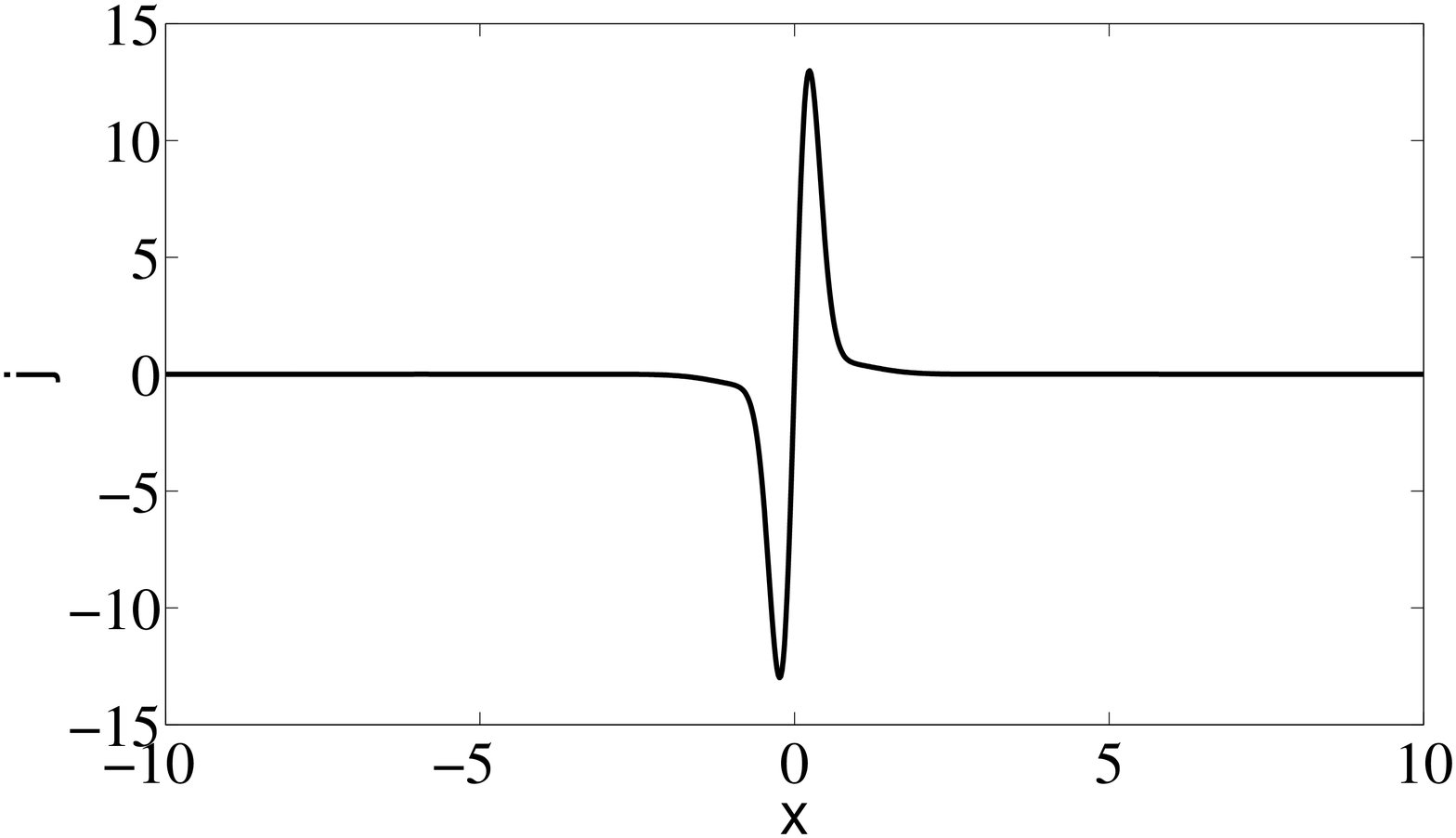}
\includegraphics[height=6.0cm,width=7.0cm]{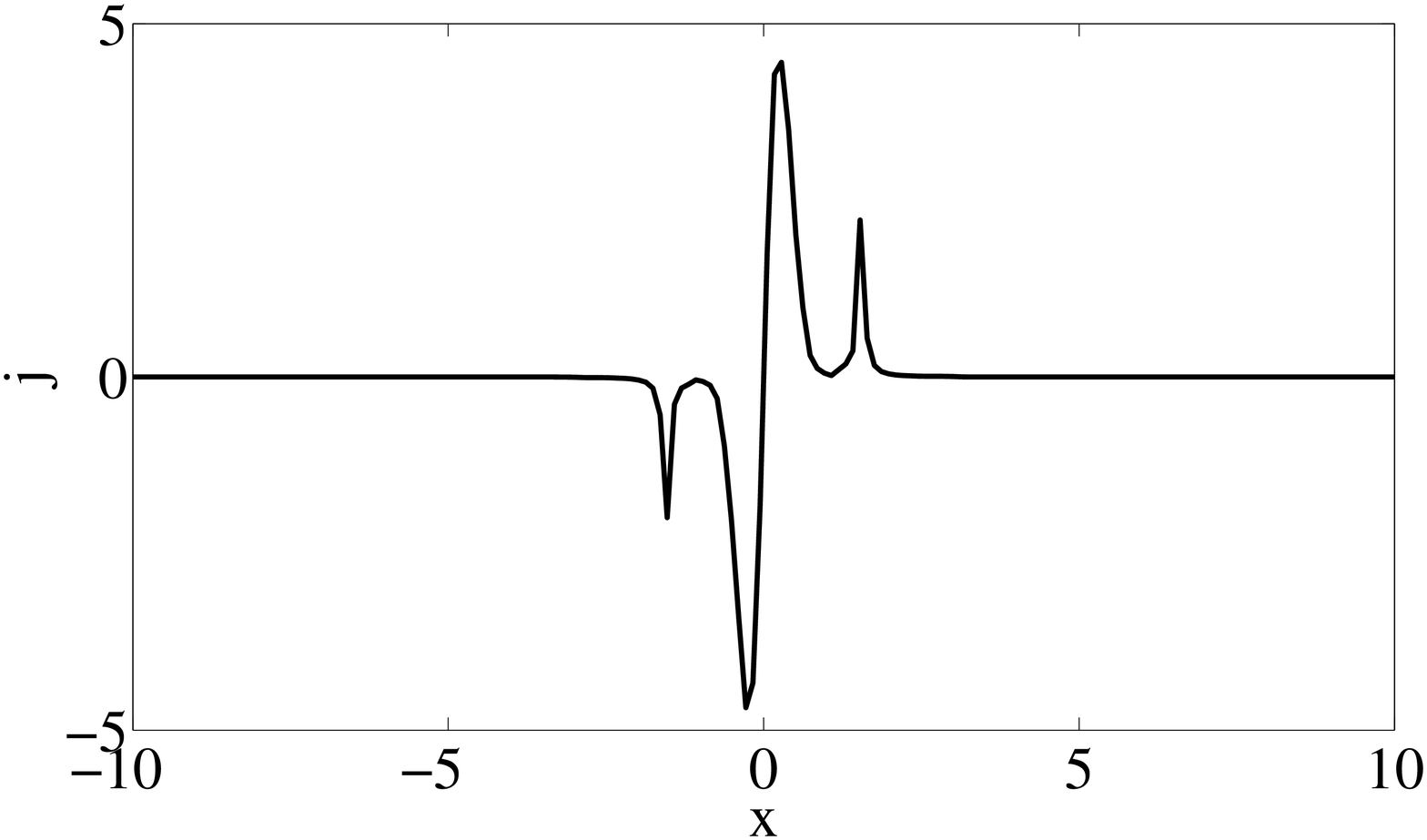}}
\caption{Density flux for $\alpha=16$. Left figure: Curve for instant of time $t=0$;
 Right figure: Curve for instant of time $t=1$.}
\label{jalpha16t1}
\end{figure}

\section{Summary and outlook}

In this paper we have presented a numerical solution of a non-Fickian
diffusion equation which is a partial differential equation of the
hyperbolic type. This equation is of physical interest in the context
of Brownian motion in inertial as well as diffusive regimes. In
our model the Brownian particle is subjected to a symmetric periodic
potential of flexible shapes (generated with a single variable parameter)
which can lead to harmonic, anharmonic or a confining potential for the
particle. 

Instead of introducing discretization in both space and time variables
we dealt with the time-derivatives through time Laplace transform and
obtained an ordinary differential equation in space variable. This
equation was then solved with a finite-difference scheme, leading to
a discretised approximate solution for $\widetilde{n}(x,s)$; the solution is
approximate due to discretization and is still formally exact in the
Laplace domain $s$. The next step consisted of numerical Laplace inversion
to obtain an approximation to the original solution $n(x,t)$.
We show that the full method is second order accurate; 
this finding receives additional support from two test examples considered
in Section 4.
One may be
able to consider further improvement. A major advantage of using the time 
Laplace transform is that we can compute the approximate solution for
long times accurately and quickly. Any iterative numerical method would
take too long to compute the solution for similar times even if we
consider an unconditionally implicit numerical method which will allow large
time steps. Additionally, our algorithm takes into consideration the
smoothness of the solution; in other words the computational effort is 
higher in the regions where the solution has steep gradients.
   Another merit of the method is that it can be easily generalized
to higher spatial dimensions. It would be of interest to
consider an application of the method to numerically solve the 
Kramers equation which is a more involved partial differential 
equation than the non-Fickian diffusion equation.

\end{document}